\documentclass{amsart}%
\usepackage{graphicx}
\usepackage{amscd}
\usepackage{amsmath}
\usepackage{amsfonts}
\usepackage{amssymb}%
\theoremstyle{plain}

\newtheorem{proposition}{Proposition}
\newtheorem{remark}{Remark}

\numberwithin{equation}{section}
\begin{document}
\title[Quantum Stochastics as a Dirac Boundary Value Problem]{Quantum Stochastics, Dirac Boundary Value Problem, and the Ultra Relativistic Limit.}
\author{V. P. Belavkin}
\address{University of Nottingham, School of Mathematics, NG7 2RD, UK}
\email{vpb@maths.nott.ac.uk}
\thanks{This work was supported by Royal Society Grant for UK - Japan collaboration.}
\thanks{Published in Rep. on Math. Phys. 46 (2000) No 3, 359 - 382}
\dedicatory{Dedicated to Roman Ingarden}\date{Received October 3, 2000.}
\maketitle

\begin{abstract}
We prove that a single-jump quantum stochastic unitary evolution is equivalent
to a Dirac boundary value problem on the half line in an extra dimension. This
amounts to the equivalence of the quantum measurement boundary-value problem
in infinite number particles space to the stochastic calculus in Fock space.
It is shown that this exactly solvable model can be obtained from a
Schr\"{o}dinger boundary value problem for a positive relativistic Hamiltonian
in the half-line as the inductive ultra relativistic limit, correspondent to
the input flow of Dirac particles with asymptotically infinite momenta. Thus
the stochastic limit can be interpreted in terms of quantum stochastic scheme
for time-continuous non-demolition observation. The question of microscopic
time reversibility is also studied for this paper.

\end{abstract}

\section{Introduction.}

All the attempts to derive the probabilistic interpretation of quantum
mechanics from classical stochastic mechanics or even from a classical chaos
(deterministic unstable dynamics) so far have been unsuccessful. The reason
for this is the nonexistence of hidden variable theories satisfying the
locality principle which can reproduce all quantum probabilities. Here we
prove an opposite point of view that any classical, as well as quantum
stochastics, can be derived from a quantum deterministic (Hamiltonian)
dynamics starting from a pure quantum state. It has been already proved in
\cite{Bel95} that the piecewise continuous stochastic unitary evolution driven
by a quantum Poisson process is equivalent to a time-dependent singular
Hamiltonian Schr\"{o}dinger problem, and the continuous stochastic unitary
evolution driven by a quantum Wiener process can be obtained as the solution
of this problem at a central limit.

There exits a broad literature on the stochastic limit in quantum physics in
which quantum stochastics is derived from a nonsingular interaction
representation of the Schr\"{o}dinger initial value problem for a quantum
field by rescaling the time and space \cite{AAFL91}. Our intention is rather
different: instead of rescaling the interaction potentials we treat the
singular interactions rigorously as the boundary conditions, and obtain the
stochastic limit as an ultra relativistic limit of the corresponding
Schr\"{o}dinger boundary value problem in a Hilbert space of infinite number
of particles. We shall prove that the discontinuous and continuous classical
as well as quantum stochastic evolutions can be obtained in this way from a
physically meaningful time continuous (in strong sense) unitary evolution by
solving a boundary value problem with an initial pure state in the extended
Hilbert space.

First we shall describe the boundary value problem corresponding to the
single-point discontinuous stochastic evolution and demonstrate the ultra
relativistic limit in this case. Then we shall obtain in a similar way the
arbitrary piece-wise continuous stochastic evolution driven by a Poisson
process starting from the second quantization of this model. And the
continuous stochastic and quantum stochastic evolution driven by a (quantum)
Wiener process is then obtained as the central limit of the strongly
continuous unitary evolution model as it was done in \cite{Bel95, Bel96} in
the singular Hamiltonian approach. But before performing this program, let us
describe the unitary toy model giving an ''unphysical'' solution of this
problem corresponding to the free hamiltonian $h\left(  p\right)  =-p$. This
toy model in the second quantization framework was suggested for the
derivation of quantum time-continuous measurement process in \cite{Bel94}.
Recently Chebotarev \cite{Che97} has shown that the secondary quantized
time-continuous toy Hamiltonian model in Fock space with a discontinuity
condition is equivalent to the Hudson-Parthasarathy (HP) quantum stochastic
evolution model \cite{HP84} in the case of commuting operator-valued
coefficients of the HP-equation. Our approach is free from the commutativity
restriction for the coefficients, and we deal with time-reversible Dirac
Hamiltonian and the boundary rather than physically meaningless discontinuity
condition and time irreversible $-p$. Moreover, we shall prove that the
stochastic model can be obtained from a positive relativistic Hamiltonian as
an inductive ultra relativistic limit on a union of Hardy class Hilbert
spaces. We call this limit the inductive stochastic approximation.

\section{A toy Hamiltonian model.}

Here we demonstrate on a toy model how the time-dependent single-point
stochastic Hamiltonian problem can be treated as an interaction representation
of a self-adjoint boundary-value Schr\"{o}dinger problem for a
strongly-continuous unitary group evolution.

Let $\mathcal{H}$ be a Hilbert space, $H$ be a bounded from below self-adjoint
operator, and $S$ be a unitary operator in $\mathcal{H}$, not necessarily
commuting with $H$. The operator $H$ called Hamiltonian, is the generator for
the conservative evolution of a quantum system, described by the
Schr\"{o}dinger equation $\mathrm{i}\hbar\partial_{t}\eta=H\eta$, and the
operator $S$ called scattering, describes the unitary quantum jump
$\eta\mapsto S\eta$ of the state vectors $\eta\in\mathcal{H}$ caused by a
singular potential interaction in the system, with the continuous unitary
evolution $\eta\mapsto e^{-\frac{\mathrm{i}}{\hbar}tH}\eta$ when there is no
jump. As for an example of such jump we can refer to the von Neumann singular
Hamiltonian model for indirect instantaneous measurement of a quantum particle
position $x\in\mathbb{R}$ via the registration of an apparatus pointer
position $y\in\mathbb{R}$. It can be described \cite{Bel95, Bel96} by the
$x$-pointwise shift $S=\hat{\sigma}$ by $\sigma\left(  x\right)
=e^{x\partial_{y}}$ of $y$ in the Hilbert space $L^{2}\left(  \mathbb{R}%
^{2}\right)  $ of square-integrable functions $\eta\left(  x,y\right)  $, and
it does not commute with the free Hamiltonian operator $H=\frac{1}{2}\left(
y^{2}-\partial_{x}^{2}\right)  $ say, of the system \textquotedblright quantum
particle plus apparatus pointer\textquotedblright.

It is usually assumed that the quantum jump occurs at a random instant of time
$t=s$ with a given probability density $\rho\left(  s\right)  >0$ on the
positive half of line $\mathbb{R}^{+}$. If $H$ and $S$ commute, the
single-point discontinuous in $t$ stochastic evolution can formally be
described by the time-dependent Schr\"{o}dinger initial value problem
\[
\mathrm{i}\hbar\partial_{t}\chi\left(  t\right)  =H_{s}\left(  t\right)
\chi\left(  t\right)  ,\quad\chi\left(  0\right)  =\eta
\]
with the singular stochastic Hamiltonian
\begin{equation}
H_{s}\left(  t\right)  =H+\mathrm{i}\hbar\delta_{s}\left(  t\right)  \ln
S,\label{1.1}%
\end{equation}
where $\delta_{s}\left(  t\right)  =\delta\left(  t-s\right)  =\delta
_{t}\left(  s\right)  $ is the Dirac $\delta$-function of $z=s-t$. Indeed,
integrating the time-dependent Hamiltonian $H_{r}\left(  s\right)  $ over $r$
from $0$ to $t$ for a fixed $s\in\mathbb{R}$, one can obtain $\chi\left(
t\right)  =V\left(  t,s\right)  \eta$ with
\[
V\left(  t,s\right)  =e^{-\frac{\mathrm{i}}{\hbar}\int_{0}^{t}H_{s}\left(
r\right)  \mathrm{d}r}=e^{-\frac{\mathrm{i}}{\hbar}tH}S^{\Delta_{0}^{t}\left(
s\right)  }=e^{\frac{\mathrm{i}}{\hbar}\left(  s-t\right)  H}S^{\Delta_{0}%
^{t}\left(  s\right)  }e^{-\frac{\mathrm{i}}{\hbar}sH},
\]
where $\Delta_{0}^{t}\left(  s\right)  =\int_{0}^{t}\delta_{r}\left(
s\right)  \mathrm{d}r$ is identified with the indicator function $1_{[0,t)}$
of the interval $[0,t)$ for a $t>0$ (at $t\leq0$ it is zero if $s>0$). The
right hand side is the form of the unitary stochastic evolution $V\left(
t,s\right)  $ which should remain valid even if the operators $H$ and $S$ do
not commute. First the evolution is conservative and continuous, $V\left(
t,s\right)  =e^{-\frac{\mathrm{i}}{\hbar}tH}$ for $t\in\lbrack0,s)$, then the
quantum jump $S$ is applied at $t=s$, and at $t>s$ the evolution is again
continuous, described by the Hamiltonian $H$. As it was noted in \cite{Bel95},
the rigorous form of the stochastic Schr\"{o}dinger equation which gives such
solution even for noncommuting $H$ and $S$ in the positive direction of $t$,
is the Ito differential equation
\begin{equation}
\mathrm{d}_{t}V\left(  t,s\right)  +\frac{\mathrm{i}}{\hbar}HV\left(
t,s\right)  \mathrm{d}t=\left(  S-I\right)  V\left(  t,s\right)
\mathrm{d}1_{t}\left(  s\right)  ,\;t>0,\quad V\left(  0,s\right)
=I.\label{1.2}%
\end{equation}
Here $\mathrm{d}_{t}V\left(  t,s\right)  =V\left(  t+\mathrm{d}t,s\right)
-V\left(  t,s\right)  $ is the forward differential corresponding to an
infinitesimal increment $\mathrm{d}t>0$ at $t$, and $\mathrm{d}1_{t}\left(
s\right)  =\Delta_{0}^{\mathrm{d}t}\left(  s-t\right)  $ is the indicator
function $\Delta_{t}^{\mathrm{d}t}\left(  s\right)  =1_{[t,t+\mathrm{d}%
t)}\left(  s\right)  $, the forward increment of the Heaviside function
$1_{t}\left(  s\right)  =1_{0}\left(  s-t\right)  $, where $1_{0}%
=1_{(-\infty,0)}$. The equation (\ref{1.2}) simply means that $t\mapsto
V\left(  t\right)  $ for a fixed $s=z$ satisfies the usual Schr\"{o}dinger
equation $\mathrm{i}\hbar\partial_{t}V\left(  t\right)  =HV\left(  t\right)  $
if $t\neq s$ as $\mathrm{d}1_{t}\left(  s\right)  =0$ for a sufficiently small
$\mathrm{d}t$ ($\mathrm{d}t<s-t$ if $t<s$, and any $\mathrm{d}t>0$ if $t>s$),
while it jumps, $\mathrm{d}_{t}V=\left(  S-I\right)  V$ at $t=s$ as
$\mathrm{d}1_{t}\left(  s\right)  |_{t=s}=1\gg\mathrm{d}t.$ Integrating
$\mathrm{d}_{z}\chi\left(  z\right)  =\mathrm{d}_{z}V\left(  z\right)  \eta$
on the domain of the operator $H$ first from $0$ to $z=s$ with an initial
condition $\chi\left(  0\right)  =\eta$, and then from $s$ to $t$ with the
initial condition $\chi\left(  s_{+}\right)  :=\lim_{z\searrow s}\chi\left(
z\right)  =S\chi\left(  s\right)  $ one can easily obtain the solution in the
form $\chi\left(  t,s\right)  =V\left(  t,s\right)  \eta$, where
\begin{equation}
V\left(  t,z\right)  =e^{-\frac{\mathrm{i}}{\hbar}tH}S\left(  z\right)
^{1_{[0,t)}\left(  z\right)  },\quad S\left(  z\right)  =e^{\frac{\mathrm{i}%
}{\hbar}zH}Se^{-\frac{\mathrm{i}}{\hbar}zH}\label{1.6}%
\end{equation}
without the commutativity condition for $H$ and $S$.

Now we shall prove that the stochstic single-jump discontinuous evolution
$\hat{V}\left(  t\right)  =V\left(  t,\cdot\right)  $ can be treated as the
interaction representation
\[
\left(  \hat{V}\left(  t\right)  \chi^{0}\right)  \left(  z\right)  =V\left(
t,z\right)  \chi^{0}\left(  z\right)  =\left(  e^{\frac{\mathrm{i}}{\hbar
}th\left(  \hat{p}\right)  }\chi^{t}\right)  \left(  z\right)
\]
for a deterministic strongly-continuous unitary group evolution $\chi
^{0}\mapsto\chi^{t}$ in one extra dimension $z\in\mathbb{R}$ with the initial
conditions localized at $z>0$ as $\chi^{0}\left(  z\right)  =\eta
\in\mathcal{H}$, $\chi^{0}\left(  z\right)  =0$ at $z\leq0$ such that%
\[
\int_{0}^{\infty}\left\Vert AV\left(  t,s\right)  \eta\right\Vert ^{2}%
\rho\left(  s\right)  \mathrm{d}s=\left\Vert \left(  A\otimes\hat{1}\right)
\chi^{t}\right\Vert _{\rho}^{2}.
\]
Here $A$ is any operator on $\mathcal{H}$, $h\left(  \hat{p}\right)  =-\hat
{p}$ is free Hamiltonian given by the momentum $\hat{p}=-\mathrm{i}%
\hbar\partial_{z}$ in the direction of $z\in\mathbb{R}$, and $\hat{1}$ is the
identity in $L^{2}\left(  \mathbb{R},\rho\right)  $.

\begin{proposition}
Let $\mathbb{R}\ni z\rightarrow\rho\left(  z\right)  $ be a smooth positive
symmetric function normalized as $\int_{0}^{\infty}\rho\left(  s\right)
\mathrm{d}s=1$, $L^{2}\left(  \mathbb{R},\rho\right)  $ be the space of $\rho
$-square integrable functions $g:\mathbb{R}\rightarrow\mathbb{C}$,
\[
\left\langle g|g\right\rangle _{\rho}=\int_{-\infty}^{\infty}\left\vert
g\left(  z\right)  \right\vert ^{2}\rho\left(  z\right)  \mathrm{d}%
z\equiv\left\Vert g\right\Vert _{\rho}^{2}<\infty,
\]
and $u\left(  z\right)  $\ be a locally integrable complex function with
$2\operatorname{Im}u\left(  z\right)  =\hbar\partial_{z}\ln\rho\left(
z\right)  $. Then the described stochastic Hamiltonian problem (\ref{1.2}) is
unitary equivalent to the self-adjoint boundary-value Schr\"{o}dinger problem
\begin{equation}
\mathrm{i}\hbar\partial_{t}\chi^{t}\left(  z\right)  =\left(  u\left(
z\right)  +\mathrm{i}\hbar\partial_{z}+H\right)  \chi^{t}\left(  z\right)
,\quad\chi^{t}\left(  0_{-}\right)  =S\chi^{t}\left(  0\right)
,\;t>0\label{1.3}%
\end{equation}
in the Hilbert product $\mathcal{H}\otimes L^{2}\left(  \mathbb{R}%
,\rho\right)  $ in the following sense: the unitary operators $V^{t}%
=e^{t\partial_{z}}\hat{V}\left(  t\right)  $ defined by the stochastic
evolution at $t>0$ form a strongly continuous group with $V^{-t}=V^{t\dagger}%
$, $V^{0}=I$, resolving the boundary value problem (\ref{1.3}) as $\chi
^{t}=V^{t}\chi^{0}$, $\forall\chi^{0}\in\mathcal{H}\otimes L^{2}\left(
\mathbb{R},\rho\right)  $.
\end{proposition}

\begin{proof}
The boundary value problem (\ref{1.3}) is well defined on the space of smooth
(at $z\neq0$) $\rho$-square integrable functions $\chi$ with values in a dense
$S$-invariant domain of $H$. It is symmetric as $H$ is self-adjoint, and due
to the unitary boundary and logarithmic derivative conditions
\begin{align*}
0 &  =\left(  \left\Vert S\chi\left(  0\right)  \right\Vert ^{2}-\left\Vert
\chi\left(  0\right)  \right\Vert ^{2}\right)  \rho\left(  0\right)  =\left(
\left\Vert \chi\left(  0_{-}\right)  \right\Vert ^{2}-\left\Vert \chi\left(
0\right)  \right\Vert ^{2}\right)  \rho\left(  0\right)  \\
&  =\int\left[  \partial_{z}\left(  \left\Vert \chi\left(  z\right)
\right\Vert ^{2}\rho\left(  z\right)  \right)  +\left\Vert \chi\left(
z\right)  \right\Vert ^{2}\left(  \frac{2}{\hbar}\operatorname{Im}u\left(
z\right)  \rho\left(  z\right)  -\partial_{z}\rho\left(  z\right)  \right)
\right]  \mathrm{d}z\\
&  =\frac{2}{\hbar}\int\left(  \operatorname{Im}u\left(  z\right)  \left\Vert
\chi\left(  z\right)  \right\Vert ^{2}+\hbar\operatorname{Re}\left\langle
\chi\left(  z\right)  |\chi^{\prime}\left(  z\right)  \right\rangle \right)
\rho\left(  z\right)  \mathrm{d}z=\frac{2}{\hbar}\operatorname{Im}\left\langle
\chi|\hat{h}\chi\right\rangle _{\rho}.
\end{align*}
In fact, this problem is self-adjoint on the natural extension of this domain
to the absolutely-continuous at $z\neq0$ right-continuous at $z=0$
$\mathcal{H}$-valued functions satisfying the boundary condition with the
derivatives in $\mathcal{H}\otimes L^{2}\left(  \mathbb{R},\rho\right)  $ as
it has apparently the differentiable unitary solution
\begin{equation}
\chi^{t}\left(  z\right)  =e^{\frac{\mathrm{i}}{\hbar}\int_{0}^{z}\left(
u\left(  r\right)  +H\right)  \mathrm{d}r}\chi_{t}\left(  z+t\right)
,\quad\chi_{t}\left(  s\right)  =S^{\Delta_{0}^{t}\left(  s\right)  }\chi
_{0}\left(  s\right)  ,\label{1.4}%
\end{equation}
where $\chi_{0}\left(  s\right)  =\exp\left[  -\frac{\mathrm{i}}{\hbar}%
\int_{0}^{s}\left(  u\left(  z\right)  +H\right)  \mathrm{d}z\right]  \chi
^{0}\left(  s\right)  $ for any $\chi^{0}$ from the extended domain. Indeed,
substituting $\chi^{t}=\exp\left[  \frac{\mathrm{i}}{\hbar}\int_{0}^{z}\left(
u\left(  r\right)  +H\right)  \mathrm{d}r\right]  \chi_{0}^{t}$ into the
equation \{\ref{1.3}\} we obtain the transport equation $\partial_{t}\chi
_{0}^{t}\left(  z\right)  =\partial_{z}\chi_{0}^{t}\left(  z\right)  $ with
the same boundary condition $\chi_{0}^{t}\left(  -0\right)  =S\chi_{0}%
^{t}\left(  0\right)  $ and the initial condition $\chi_{0}^{0}=\chi_{0}$
corresponding to a $\chi^{0}\in\mathcal{H}\otimes L^{2}\left(  \mathbb{R}%
,\rho\right)  $. This simple initial boundary value problem has the obvious
solution $\chi_{0}^{t}\left(  z\right)  =\chi_{t}\left(  z+t\right)  $ with
$\chi_{t}$ given in (\ref{1.4}) as
\begin{equation}
\chi_{t}\left(  s\right)  =S^{1_{(0,t]}\left(  t-s\right)  }\chi^{0}\left(
s\right)  ,\;t>0,\quad\chi_{t}\left(  s\right)  =S^{-1_{[-t,0)}\left(
s\right)  }\chi^{0}\left(  s\right)  ,\;t<0.\label{1.5}%
\end{equation}

The unitarity of $S^{\Delta_{0}^{t}\left(  s\right)  }$ in $\mathcal{H}$ and
of shift $e^{t\partial_{z}}$ in $L^{2}\left(  \mathbb{R}\right)  $ implies the
unitarity of the resolving map $V^{t}:\chi^{0}\mapsto\chi^{t}$ in
$\mathcal{H}\otimes L^{2}\left(  \mathbb{R},\rho\right)  $,
\[
\left\Vert \chi^{t}\right\Vert _{\rho}^{2}=\left\Vert \chi_{t}\right\Vert
^{2}\rho\left(  0\right)  =\left\Vert \chi_{0}\right\Vert ^{2}\rho\left(
0\right)  =\left\Vert \chi^{0}\right\Vert _{\rho}^{2}%
\]
because $\rho\left(  z\right)  =\rho\left(  0\right)  \exp\left[  \frac
{1}{\hbar}\int_{0}^{z}2\operatorname{Im}u\left(  r\right)  \mathrm{d}r\right]
$ and
\begin{align*}
\left\Vert \chi\right\Vert _{\rho}^{2} &  =\int\left\Vert e^{\frac{\mathrm{i}%
}{\hbar}\int_{0}^{z}\left(  u\left(  r\right)  +H\right)  \mathrm{d}r}\chi
_{0}\left(  z\right)  \right\Vert ^{2}\rho\left(  z\right)  \mathrm{d}z\\
&  =\int\left\Vert \chi_{0}\left(  z\right)  \right\Vert ^{2}\rho\left(
0\right)  \mathrm{d}z=\left\Vert \chi_{0}\right\Vert ^{2}\rho\left(  0\right)
.
\end{align*}
Moreover, the map $t\mapsto V^{t}$ has the multiplicative representation
property $V^{r}V^{t}=V^{r+t}$ of the group $\mathbb{R}\ni r,t$ because the map
$t\mapsto S^{\Delta_{0}^{t}\left(  s\right)  }$ is a multiplicative
shift-cocycle,
\[
S^{\Delta_{0}^{r}\left(  s\right)  }e^{t\partial_{s}}S^{\Delta_{0}^{t}\left(
s\right)  }=e^{t\partial_{s}}S^{\Delta_{0}^{r+t}\left(  s\right)  }%
,\quad\forall r,t\in\mathbb{R}%
\]
by virtue of the additive cocycle property for the commuting $\Delta_{0}%
^{t}\left(  s\right)  =1_{t}\left(  s\right)  -1_{0}\left(  s\right)  $:
\[
\left[  \Delta_{0}^{r}+e^{t\partial_{s}}\Delta_{0}^{t}\right]  \left(
s\right)  =1_{r}\left(  s\right)  -1_{0}\left(  s\right)  +1_{t}\left(
s+t\right)  -1_{0}\left(  s+t\right)  =e^{t\partial_{s}}\Delta_{0}%
^{r+t}\left(  s\right)  .
\]

The subtraction $\chi\left(  t,z\right)  =e^{\frac{\mathrm{i}}{\hbar}t\hat{h}%
}\chi^{t}\left(  z\right)  $ of free evolution with the generator $\hat
{h}g\left(  z\right)  =\left(  u\left(  z\right)  +\mathrm{i}\hbar\partial
_{z}\right)  g\left(  z\right)  $ obviously gives
\begin{align*}
\chi\left(  t,s\right)   &  =e^{\frac{\mathrm{i}}{\hbar}\int_{0}^{t}u\left(
s-r\right)  \mathrm{d}r}\chi^{t}\left(  s-t\right)  =e^{\frac{\mathrm{i}%
}{\hbar}\left(  \int_{0}^{s}u\left(  s-r\right)  \mathrm{d}r+\left(
s-t\right)  H\right)  }\chi_{t}\left(  s\right)  \\
&  =e^{\frac{\mathrm{i}}{\hbar}\left(  \int_{0}^{s}u\left(  z\right)
\mathrm{d}z+\left(  s-t\right)  H\right)  }S^{\Delta_{0}^{t}\left(  s\right)
}e^{-\frac{\mathrm{i}}{\hbar}\int_{0}^{s}\left(  u\left(  z\right)  +H\right)
\mathrm{d}z}\chi^{0}=V\left(  t,s\right)  \chi^{0}\left(  s\right)  ,
\end{align*}
Thus the single-point discontinuous unitary $e^{-\frac{\mathrm{i}}{\hbar}%
t\hat{h}}$-cocycle
\[
V\left(  t,s\right)  =e^{\frac{\mathrm{i}}{\hbar}t\hat{h}}V^{t}=e^{\frac
{\mathrm{i}}{\hbar}\left(  s-t\right)  H}S^{\Delta_{0}^{t}\left(  s\right)
}e^{-\frac{\mathrm{i}}{\hbar}sH},\quad t\in\mathbb{R}%
\]
with $\Delta_{0}^{t}\left(  s\right)  =1_{[0,t)}\left(  s\right)  $ for a
positive $t$ and $s\in\mathbb{R}^{+}$, solves indeed the single-jump Ito
equation (\ref{1.2}). It describes the interaction representation for the
strongly continuous unitary group evolution $V^{t}$ resolving the boundary
value problem (\ref{1.3}) with initially constant functions $\chi^{0}\left(
s\right)  =\eta$ at $s>0.$ Moreover,
\[
\left\Vert \left(  A\otimes\hat{1}\right)  V^{t}\chi^{0}\right\Vert _{\rho
}^{2}=\left\Vert \left(  A\otimes\hat{1}\right)  \chi\left(  t\right)
\right\Vert _{\rho}^{2}%
\]
coinsides with the expectation of $\left\Vert AV\left(  t,\cdot\right)
\eta\right\Vert ^{2}$ for any $\eta\in\mathcal{H}$ and $A\in\mathcal{B}\left(
\mathcal{H}\right)  $ if $\chi^{0}\left(  z\right)  =0$ at $z\leq0$ because
$\chi\left(  t,z\right)  =V\left(  t,z\right)  \chi^{0}\left(  z\right)  =0$
at $z\leq0$. Note that such localized initial $\chi^{0}\in\mathcal{H}\otimes
L^{2}\left(  \mathbb{R},\rho\right)  $ are absolutely continuous with zero
derivative at $z\neq0$, and they satisfy the boundary condition: $\chi
^{0}\left(  -0\right)  =0=S\chi^{0}\left(  0\right)  $, but they are not
strictly speaking in the domain of the self-adjoint generator for $V^{t}$ as
they are not right continuoous at $z=0$.
\end{proof}

\begin{remark}
The toy Schr\"{o}dinger boundary value problem (\ref{1.3}) is unphysical in
three aspects. First, the equation (\ref{1.3}) is not invariant under the
reversion of time arrow, i.e. under an isometric complex conjugation
$\eta\mapsto\bar{\eta}$ and the reflection $t\mapsto-t$, even if $\bar
{S}=S^{-1}$ and $\operatorname{Im}H=0$ as the Hamiltonian $\hat{h}=\hat
{u}+\mathrm{i}\hbar\partial_{z}$ is not real, $\operatorname{Im}\hat
{h}=\operatorname{Im}\hat{u}+\hbar\partial_{z}$. Second, a physical wave
function $\psi^{t}\left(  z\right)  $ should have a continuous propagation in
both directions of $z$, and at the boundary must have a jump not in the
coordinate but in momentum representation. The momentum can change its
direction but not the magnitude (conservation of momentum) in the result of
the singular interaction with the boundary. And third, the free Hamiltonian
$\hat{h}$ must be bounded from below which is not so in the case of
hamiltonian function $h\left(  z,p\right)  =u\left(  z\right)  -p$
corresponding to the equation (\ref{1.3}).
\end{remark}

Now we show how to rectify the first two failures of the toy model, but the
third, which is a more serious failure, will be sorted out in the next
sections by considering the toy model as a dressed limiting case.

Instead of the single wave function $\chi^{t}\left(  z\right)  $ on
$\mathbb{R}$ let us considering the pair $\left(  \psi,\tilde{\psi}\right)  $
of input and output wave functions with
\[
\psi^{t}\left(  z\right)  =\chi^{t}\left(  z\right)  ,\;z\geq0,\quad
\tilde{\psi}^{t}\left(  -z\right)  =\chi^{t}\left(  z\right)  ,\;z<0
\]
on the half of line $\mathbb{R}^{+}$, having the scalar product
\[
\int_{0}^{\infty}\left(  \left\Vert \psi\left(  z\right)  \right\Vert
^{2}+\left\Vert \tilde{\psi}\left(  z\right)  \right\Vert ^{2}\right)
\rho\left(  z\right)  \mathrm{d}z=\int_{-\infty}^{\infty}\left\Vert
\chi\left(  z\right)  \right\Vert ^{2}\rho\left(  z\right)  \mathrm{d}%
z\text{.}%
\]
They satisfy the system of equations
\begin{align*}
\mathrm{i}\hbar\partial_{t}\psi^{t}\left(  z\right)   &  =\left(  u\left(
z\right)  +\mathrm{i}\hbar\partial_{z}+H\right)  \psi^{t}\left(  z\right)
,\quad\psi^{0}\in\mathcal{H}\otimes L^{2}\left(  \mathbb{R}^{+},\rho\right)
\\
\mathrm{i}\hbar\partial_{t}\tilde{\psi}^{t}\left(  z\right)   &  =\left(
\tilde{u}\left(  z\right)  -\mathrm{i}\hbar\partial_{z}+H\right)  \tilde{\psi
}^{t}\left(  z\right)  ,\quad\tilde{\psi}^{0}\in\mathcal{H}\otimes
L^{2}\left(  \mathbb{R}^{+},\rho\right)
\end{align*}
for a quantum system interacting with a massless Dirac particle in a static
field $\left(  u,\tilde{u}\right)  $ on $\mathbb{R}^{+}$ through the boundary
condition $\tilde{\psi}^{t}\left(  0\right)  =S\psi^{t}\left(  0\right)  $,
where $\tilde{\psi}^{t}\left(  0\right)  =\chi^{t}\left(  0_{-}\right)  $. One
can show that this is indeed the diagonal form of the Dirac equation in one
dimension in the eigen-representation of the Dirac velocity $c=-\sigma_{z}$
along $z\in\mathbb{R}^{+}$, with the electric and magnetic field components
$u_{\pm}$, given by the symmetric and antisymmetric parts $u\pm\tilde{u}$ of
$u$ on $\mathbb{R}$ in the case $\operatorname{Im}u=0$. The components of
$\left(  \psi,\tilde{\psi}\right)  $ propagate independently at $z>0$ as plane
waves in the opposite directions with a spin (or polarization) oriented in the
direction of $z$, and in the scalar case $\mathcal{H}=\mathbb{C}$ are
connected by the Dirac type boundary condition $\left(  1+\mathrm{i}%
\mu\right)  \tilde{\psi}^{t}\left(  0\right)  =\left(  1-\mathrm{i}\mu\right)
\psi^{t}\left(  0\right)  $ correspondent to a point mass $\hbar\mu$ at $z=0$.
The input wave function
\begin{equation}
\psi^{t}\left(  z\right)  =e^{-\frac{\mathrm{i}}{\hbar}\int_{0}^{t}\left(
u\left(  z+r\right)  +H\right)  \mathrm{d}r}\psi^{0}\left(  z+t\right)
=e^{-\frac{\mathrm{i}}{\hbar}t\left(  \hat{h}+H\right)  }\psi^{0}\left(
z\right)  \label{1.7}%
\end{equation}
is the solution to the equation (\ref{1.3}) at $z\in\mathbb{R}^{+}$with
$\chi^{0}|_{z>0}=\psi^{0}$ which does not need the boundary condition at $z=0$
when solving the initial value problem in $t>0$. The output wave function
satisfies the reflected equation at $z>0$ and the unitary boundary condition
at $z=0$:
\begin{equation}
\mathrm{i}\hbar\partial_{t}\tilde{\psi}^{t}\left(  z\right)  =\left(
\tilde{u}\left(  z\right)  -\mathrm{i}\hbar\partial_{z}+H\right)  \tilde{\psi
}^{t}\left(  z\right)  ,\quad\tilde{\psi}^{t}\left(  0\right)  =S\psi
^{t}\left(  0\right)  ,\label{1.8}%
\end{equation}
where $\tilde{u}$ $\left(  z\right)  =u\left(  -z\right)  $. It has the
solution
\[
\tilde{\psi}^{t}\left(  z\right)  =e^{-\frac{\mathrm{i}}{\hbar}\int_{0}%
^{t}\left(  \tilde{u}\left(  z-r\right)  +H\right)  \mathrm{d}r}\left[
\tilde{\psi}^{0}\left(  z-t\right)  1_{t}^{\bot}\left(  z\right)  +S\left(
t-z\right)  \psi^{0}\left(  t-z\right)  1_{t}\left(  z\right)  \right]  ,
\]
where $1_{t}^{\bot}\left(  z\right)  =1-1_{t}\left(  z\right)  $. This can be
written in the similar way as $\psi^{t}$,
\begin{equation}
\tilde{\psi}^{t}\left(  z\right)  =e^{-\frac{\mathrm{i}}{\hbar}\int_{0}%
^{t}\left(  \tilde{u}\left(  z-r\right)  +H\right)  \mathrm{d}r}\tilde{\psi
}^{0}\left(  z-t\right)  =e^{-\frac{\mathrm{i}}{\hbar}t\left(  \check
{h}+H\right)  }\tilde{\psi}^{0}\left(  z\right)  \label{1.9}%
\end{equation}
with $\check{h}=\tilde{u}\left(  z\right)  -\mathrm{i}\hbar\partial_{z}$ if
$\psi^{0}\left(  z\right)  $ is extended into the domain $z<0$ as
\begin{equation}
S\left(  z\right)  \psi^{0}\left(  z\right)  =\tilde{\psi}^{0}\left(
-z\right)  ,\quad S\left(  z\right)  =e^{\frac{\mathrm{i}}{\hbar}zH}%
Se^{-\frac{\mathrm{i}}{\hbar}zH}\label{1.10}%
\end{equation}
Note that reflection condition (\ref{1.10}) remains valid for all $t>0$ if
$\psi^{t}$ is extended into the region $z<0$ by the solution of (\ref{1.7})
for any $t\in\mathbb{R}^{+}$:
\begin{align*}
\tilde{\psi}^{t}\left(  -z\right)   &  =e^{-\frac{\mathrm{i}}{\hbar}\int
_{0}^{t}\left(  u\left(  z+r\right)  +H\right)  \mathrm{d}r}S\left(
t+z\right)  \psi^{0}\left(  t+z\right)  \\
&  =S\left(  z\right)  e^{-\frac{\mathrm{i}}{\hbar}\int_{0}^{t}\left(
u\left(  z+r\right)  +H\right)  \mathrm{d}r}\psi^{0}\left(  z+t\right)
=S\left(  z\right)  \psi^{t}\left(  z\right)  .
\end{align*}
Extending also the output wave $\tilde{\psi}^{t}$ by (\ref{1.9}) into the
region $z<0$, where $\tilde{\psi}^{0}\left(  z\right)  $ is extended into
$z<0$ by (\ref{1.10}), we obtain the continuous propagation of $\psi
,\tilde{\psi}$ through the boundary in the opposite directions, with the
unitary reflection connection (\ref{1.10}) for all $z\in\mathbb{R}$. If
$\operatorname{Re}u\left(  z\right)  $ is symmetric (no magnetic field) and
$\bar{H}=H$, where $\bar{H}\eta=\overline{H\bar{\eta}}$ with respect to a
complex conjugation in $\mathcal{H}$, then the system of Schr\"{o}dinger
equations for the pair $\left(  \psi,\tilde{\psi}\right)  $ remains invariant
under the time reflection with complex conjugation up to exchange $\bar{\psi
}^{-t}\rightleftarrows\tilde{\psi}^{t}$. Indeed, as $\operatorname{Im}u\left(
z\right)  =\hbar\partial_{z}\ln\sqrt{\rho\left(  z\right)  }$ is atisymmetric,
in this case $\tilde{u}\left(  z\right)  =u\left(  -z\right)  =\bar{u}\left(
z\right)  $, and the complex conjugated hamiltonian $\overline{\hat{h}}%
=\bar{u}\left(  z\right)  -\mathrm{i}\hbar\partial_{z}$ coinsides with the
operator $\check{h}$ corresponding to $\tilde{h}\left(  z,p\right)  =\bar
{u}\left(  z\right)  +p=\bar{h}\left(  z,p\right)  $. The boundary value
problem is invariant under time reversion if $\bar{S}=S^{-1}$ as the
reflection condition (\ref{1.10}) is extended to the negative $t$ by the
exchange due to $S\left(  z\right)  ^{-1}=\bar{S}\left(  -z\right)  .$ Thus
the reversion of time arrow is equivalent to the exchange of the input and
output wave functions which is an involute isomorphism due to
\[
\int_{-\infty}^{\infty}\left\Vert \psi\left(  z\right)  \right\Vert ^{2}%
\rho\left(  z\right)  \mathrm{d}z=\left\Vert \chi\right\Vert _{\rho}^{2}%
=\int_{-\infty}^{\infty}\left\Vert \tilde{\psi}\left(  z\right)  \right\Vert
^{2}\rho\left(  z\right)  \mathrm{d}z.
\]

\section{A unitary reflection model}

As we have seen in the end of the previous section, a unitary quantum state
jump at a random instant of time $s\geq0$ is a result of solving of the toy
Schr\"{o}dinger boundary value problem in the interaction representation for a
strongly continuous unitary evolution of a Dirac particle with zero mass. The
input particle, an ''instanton'' with the state vectors defining the input
probabilities for $s=z$, has the unbounded from below kinetic energy $e\left(
p\right)  =-p$ corresponding to the constant negative velocity $v=e^{\prime
}\left(  p\right)  =-1$ along the intrinsic time coordinate $z$ which does not
coincide with the direction of the momentum if $p>0$. One can interpret such
strange particle as a trigger for instantaneous measurement in a quantum
system at the time $z\in\mathbb{R}^{+}$, and might like to consider it as a
normal particle, like a ''bubble'' in a cloud chamber on the boundary of
$\mathbb{R}^{d}\times\mathbb{R}^{+}$ as it was assumed in \cite{Bel95}, with
positive kinetic energy and a non-zero mass.

Our aim is to obtain the instanton as an ultra relativistic limit of a quantum
particle with a positive kinetic energy corresponding to a mass $m_{0}\geq0$.
Here we shall treat the kinetic energy separately for input and output
instantons as a function of the momentum $p\in\mathbb{R}^{-}$ and
$p\in\mathbb{R}^{+}$ respectively along a coordinate $z\in\mathbb{R}^{+}$ with
the same self-adjoint operator values $e\left(  p\right)  \geq0$ in a Hilbert
space $\mathfrak{h}$ of its spin or other degrees of freedom.

For example one can take the relativistic mass operator-function
\begin{equation}
e\left(  p\right)  =\left(  p^{2}+\hbar^{2}\mu^{2}\right)  ^{1/2},\quad\mu
^{2}=\mu_{0}^{2}-\triangledown^{2} \label{2.1}%
\end{equation}
in the Hilbert space $\mathfrak{h}=L^{2}\left(  \mathbb{R}^{d}\right)  $ which
defines the velocities $v\left(  p\right)  =p/e\left(  p\right)  =e^{\prime
}\left(  p\right)  $ with the same signature as $p$.

At the boundary $z=0$ the incoming particle with the negative momentum $p<0$
is reflected into the outgoing one with the opposite momentum $-p$. The
singular interaction with the boundary causes also a quantum jump in other
degrees of freedom. It is described by the unitary operator $\sigma$ in
$\mathfrak{h}$ which is assumed to commute with $e\left(  p\right)  $ for each
$p$ as it is in the quantum measurement model \cite{Bel95} when $\sigma
=e^{\mathbf{x}\partial_{\mathbf{y}}}$ with $\nabla=\partial_{\mathbf{y}}$ in
(\ref{2.1}).

Let $\mathfrak{h}$ be a Hilbert space with isometric complex conjugation
$\mathfrak{h}\ni\eta\mapsto\bar{\eta}\in\mathfrak{h}$, and $L_{\mathfrak{h}%
}^{2}\left(  \mathbb{R}^{-}\right)  =\mathfrak{h}\otimes L^{2}\left(
\mathbb{R}^{-}\right)  $ be the space of square-integrable vector-functions
$f\left(  k\right)  \in\mathfrak{h}$ on the half-line $\mathbb{R}^{-}\ni k$.
We denote by $\mathcal{E}^{-}$ the isomorphic space of Fourier integrals
\[
\varphi\left(  z\right)  =\frac{1}{2\pi}\int_{-\infty}^{0}e^{\mathrm{i}%
kz}f\left(  k\right)  \mathrm{d}k,\quad f\in L_{\mathfrak{h}}^{2}\left(
\mathbb{R}^{-}\right)  .
\]
having the analytical continuation into the complex domain $\operatorname{Im}%
z<0$, which is called Hardy class of $\mathfrak{h}$-valued functions. One can
interpret $L_{\mathfrak{h}}^{2}\left(  \mathbb{R}^{-}\right)  $ as the Hilbert
space of quantum input states with negative momenta $p_{k}=\hbar k$, $k<0$
along $z\in\mathbb{R}$ and spin states $\eta\in\mathfrak{h}$. The generalized
eigen-functions
\begin{equation}
\varphi_{k}\left(  z\right)  =\exp\left[  \mathrm{i}kz\right]  \eta
_{k},\;k<0,\quad e\left(  \hbar k\right)  \eta_{k}=\hbar\varepsilon_{k}%
\eta_{k}\label{2.2}%
\end{equation}
corresponding to spectral values $\varepsilon_{k}\in\mathbb{R}^{+}$ of
$\varepsilon\left(  k\right)  =\hbar^{-1}e\left(  \hbar k\right)  $, are given
as the harmonic waves moving from infinity towards $z=0$ with the phase speed
$\varsigma_{k}=\varepsilon_{k}/\left\vert k\right\vert $ along $z$. The
amplitudes $\eta_{k}$ are arbitrary in $\mathfrak{h}$ if all $e\left(
p\right)  $ are proportional to the identity operator $1$ in $\mathfrak{h}$,
$\varepsilon\left(  k\right)  $ $=\varepsilon_{k}1$, as it was in the previous
section where $\mathfrak{h}=\mathcal{H}$.

The singular interaction creates the output states in the same region $z>0$ of
observation where the input field is, by the momentum inversion $p=-p_{k}%
\mapsto\tilde{p}=p_{k}$, reflecting the input wave functions $\varphi
\in\mathcal{E}^{-}$ isometrically onto
\[
\tilde{\varphi}\left(  s\right)  =\frac{1}{2\pi}\int_{-\infty}^{0}%
e^{-\mathrm{i}ks}\tilde{f}\left(  k\right)  \mathrm{d}k=\sigma\varphi\left(
-s\right)  ,\quad s\in\mathbb{R}_{+}%
\]
by $\tilde{f}\left(  k\right)  =\sigma f\left(  k\right)  ,$ $k<0$. The space
$\mathcal{E}^{+}=\left\{  \tilde{\varphi}:\varphi\in\mathcal{E}^{-}\right\}  $
is the conjugated Hardy subspace $\mathcal{E}^{+}=\left\{  \bar{\varphi
}:\varphi\in\mathcal{E}^{-}\right\}  $ of analytical functions $\bar{\varphi
}\left(  z\right)  =\overline{\varphi\left(  \bar{z}\right)  }$ in
$\operatorname{Im}z>0$. The reflected wave function satisfies the boundary
condition $\tilde{\varphi}\left(  0\right)  =\sigma\varphi\left(  0\right)  $
corresponding to the zero probability current
\[
j\left(  z\right)  =\left\Vert \tilde{\varphi}\left(  z\right)  \right\Vert
^{2}-\left\Vert \varphi\left(  z\right)  \right\Vert ^{2}%
\]
at $z=0$, and together with the input wave function $\varphi\left(  s\right)
,s\geq0$ represents the Hilbert square norms (total probability) in
$\mathcal{E}^{-}$ and $\mathcal{E}^{+}$ by the sum of the integrals over the
half-region $\mathbb{R}_{+}$:
\[
\int_{-\infty}^{\infty}\left\Vert \varphi\left(  z\right)  \right\Vert
^{2}\mathrm{d}z=\int_{0}^{\infty}\left(  \left\Vert \varphi\left(  s\right)
\right\Vert ^{2}+\left\Vert \widetilde{\varphi}\left(  s\right)  \right\Vert
^{2}\right)  \mathrm{d}s=\int_{-\infty}^{\infty}\left\Vert \tilde{\varphi
}\left(  z\right)  \right\Vert ^{2}\mathrm{d}z.
\]

As usual we shall define the free dynamics of the input and output wave
functions by the unitary propagation
\begin{align}
\varphi^{t}\left(  z\right)   &  =\frac{1}{2\pi}\int_{-\infty}^{0}%
e^{\mathrm{i}k\left(  t\varsigma\left(  k\right)  +z\right)  }f\left(
k\right)  \mathrm{d}k=\left[  e^{-\mathrm{i}t\hat{\varepsilon}}\varphi\right]
\left(  z\right)  ,\label{2.3}\\
\widetilde{\varphi}^{t}\left(  z\right)   &  =\frac{1}{2\pi}\int_{-\infty}%
^{0}e^{\mathrm{i}k\left(  t\varsigma\left(  k\right)  -z\right)  }\tilde
{f}\left(  k\right)  \mathrm{d}k=\left[  e^{-\mathrm{i}t\check{\varepsilon}%
}\widetilde{\varphi}\right]  \left(  z\right)  ,\nonumber
\end{align}
of a superposition of the harmonic eigen-functions (\ref{2.2}) in the negative
and positive direction of $z\in\mathbb{R}$ respectively with the same phase
speeds $\varsigma_{k}>1$ which are the eigen-values of the positive operators
$\varsigma\left(  k\right)  =\left\vert k\right\vert ^{-1}\varepsilon\left(
k\right)  $. The generating self-adjoint operators $\hat{\varepsilon}$,
$\check{\varepsilon}$ are the restrictions $\hat{\varepsilon}=\varepsilon
\left(  \mathrm{i}\partial_{z}\right)  |\mathcal{D}^{-}$, $\check{\varepsilon
}=\varepsilon\left(  \mathrm{i}\partial_{z}\right)  |\mathcal{D}^{+}$ of the
kinetic energy operator given by the symmetric function $\varepsilon\left(
p\right)  $ on its symmetric dense domain $\mathcal{D}\subseteq
L_{\mathfrak{h}}^{2}\left(  \mathbb{R}\right)  $, to the dense domains
$\mathcal{D}^{\mp}=\mathcal{D}\cap\mathcal{E}^{\mp}$ in the invariant
subspaces $\mathcal{E}^{\mp}\subseteq L_{\mathfrak{h}}^{2}\left(
\mathbb{R}\right)  $.

Instead of dealing with the free propagation of the input-output pair $\left(
\varphi,\tilde{\varphi}\right)  $ at the region $z>0$ with the boundary
condition $\tilde{\varphi}^{t}\left(  0\right)  =\sigma\varphi^{t}\left(
0\right)  $, it is convenient to introduce just one wave function
\[
\phi^{t}\left(  z\right)  =\varphi^{t}\left(  z\right)  ,\;\operatorname{Re}%
z\geq0,\quad\phi^{t}\left(  z\right)  =\tilde{\varphi}^{t}\left(  -z\right)
,\;\operatorname{Re}z<0
\]
considering the reflected wave as propagating in the negative direction into
the region $z<0$. Each $\phi$ coinsides with a Hardy class function
$\varphi\in\mathcal{E}^{-}$ at $z\geq0$, as well as it is a Hardy class
function at $z<0$. However $\phi\left(  -z\right)  $, $z>0$ coinsides with
another Hardy class function $\sigma\varphi\in\mathcal{E}^{-}$ such that the
continuity of the analytical functions $\varphi$ at $\operatorname{Re}z=0$
corresponds to the left discontinuity $\phi\left(  0_{-}\right)  =\sigma
\phi\left(  0\right)  $ of
\[
\phi\left(  z\right)  =1_{0}\left(  -z\right)  \varphi\left(  z\right)
+1_{0}\left(  z\right)  \sigma\varphi\left(  z\right)  ,\quad1_{0}\left(
z\right)  =\{_{1,\;z<0}^{0,\;z\geq0},
\]
where $\phi\left(  0_{-}\right)  $ is defined as the left lower sectorial
limit of $\phi\left(  z\right)  $ at $\operatorname{Re}z\nearrow0$,
$\operatorname{Im}z\nearrow0$. Obviously the Hilbert subspace $\sigma^{\hat
{1}_{0}}\mathcal{E}^{-}\subset L_{\mathfrak{h}}^{2}\left(  \mathbb{R}\right)
$ of such wave functions is isomorphic to $\mathcal{E}^{-}$ by the unitary
operator $\sigma^{\hat{1}_{0}}=I+\hat{1}_{0}\left(  \sigma-I\right)  $, where
$\hat{1}_{0}$ is the multiplication operator of $\varphi\left(  z\right)  $ by
$1$ if $z<0$, and by $0$ if $z\geq0.$ The unitary evolution group
$\upsilon^{t}=\sigma^{\hat{1}_{0}}e^{-\mathrm{i}t\hat{\varepsilon}}%
\sigma^{-\hat{1}_{0}},t\in\mathbb{R}$ for
\begin{equation}
\phi^{t}\left(  z\right)  =\varphi^{t}\left(  z\right)  +1_{0}\left(
z\right)  \left(  \sigma-1\right)  \varphi^{t}\left(  z\right)  =\sigma
^{1_{0}\left(  z\right)  }\varphi^{t}\left(  z\right)  ,\label{2.4}%
\end{equation}
is unitary equivalent but different from the free propagation $e^{-\mathrm{i}%
t\hat{\varepsilon}}$ of $\varphi^{t}$ in $\mathbb{R}$. Each harmonic
eigen-function (\ref{2.2}) having the plane wave propagation
\[
\varphi_{k}^{t}\left(  z\right)  =e^{-\mathrm{i}\varepsilon_{k}t}\varphi
_{k}\left(  z\right)  =\varphi_{k}\left(  z+\varsigma_{k}t\right)  ,
\]
for the negative $k\in\mathbb{R}^{-}$, \ is now truncated, $\phi_{k}\left(
z\right)  =e^{\mathrm{i}kz}\sigma^{1_{0}\left(  z\right)  }\eta_{k}$, and
propagates in the negative direction as
\[
\phi_{k}^{t}\left(  z\right)  =\sigma^{1_{0}\left(  z\right)  }\varphi
_{k}\left(  z+\varsigma_{k}t\right)  =e^{-\mathrm{i}\varepsilon_{k}t}\phi
_{k}\left(  z\right)  \neq\phi_{k}\left(  z+\varsigma_{k}t\right)  ,
\]
keeping the truncation at $z=0$. Therefore the subtraction $\phi_{t}\left(
z\right)  =e^{\mathrm{i}t\hat{\varepsilon}}\phi^{t}\left(  z\right)  $ of the
free propagation of $\varphi^{t}$ from $\phi^{t}$ does not return it to the
initial $\phi^{0}=\sigma^{\hat{1}_{0}}\varphi^{0}$ but to $\phi_{t}%
=\sigma^{\hat{\pi}^{t}}\varphi^{0}=\upsilon_{t}\phi_{0}$, where $\hat{\pi}%
^{t}=e^{\mathrm{i}t\hat{\varepsilon}}\hat{1}_{0}e^{-\mathrm{i}t\hat
{\varepsilon}}$, $\upsilon_{t}=\sigma^{\hat{\pi}^{t}}\sigma^{-\hat{1}_{0}}$,
and $\phi_{0}=\phi^{0}$. Thus we have proved the following proposition for the
particular case $\varkappa=0$ of an operator-function $\varkappa$ on
$\mathbb{R}$, defined in the Proposition 1 of previous section as
$\hbar\varkappa\left(  z\right)  =u\left(  z\right)  I+H$ in $\mathfrak{h}%
=\mathcal{H}$.

Let $\varkappa\left(  z\right)  $, $z\in\mathbb{R}$ be an operator-valued
function in $\mathfrak{h}$ defining a smooth symmetric density function $\rho$
on $\mathbb{R}$ with values $\rho\left(  z\right)  \in\mathcal{L}\left(
\mathfrak{h}\right)  $ in the continuous operators by
\[
\mathrm{i}\partial_{z}\rho\left(  z\right)  =\rho\left(  z\right)
\varkappa\left(  z\right)  -\varkappa\left(  z\right)  ^{+}\rho\left(
z\right)  =0,\quad\rho\left(  0\right)  =\rho_{0},
\]
where $\varkappa\left(  z\right)  ^{+}$ is the Hermitian adjoint in
$\mathfrak{h}$, and $\rho_{0}$ is a positive invertible operator
$\mathfrak{h}\rightarrow\mathfrak{h}$. The function $\varkappa$ is assumed to
be locally integrable in the sense that it generates the one-parametric
exponential family
\begin{equation}
\epsilon_{\varkappa}\left(  \pm z\right)  =\overrightarrow{\exp}\left[
-\mathrm{i}\int_{0}^{\pm z}\varkappa\left(  \pm s\right)  \mathrm{d}s\right]
,\quad z>0\label{2.5}%
\end{equation}
as a solution to the equation $\mathrm{i}\partial_{z}\epsilon_{\varkappa
}=\epsilon_{\varkappa}\varkappa$ in both directions of $z$ with the boundary
condition $\epsilon_{\varkappa}\left(  0\right)  =1$ such that $\rho\left(
z\right)  =\epsilon_{\varkappa}\left(  z\right)  ^{+}\rho_{0}\epsilon
_{\varkappa}\left(  z\right)  $. Below we shall denote by $\hat{\epsilon
}_{\varkappa}$ and $\check{\epsilon}_{\varkappa}$ the operators of pointwise
multiplication by the functions $\epsilon_{\varkappa}:z\mapsto\epsilon
_{\varkappa}\left(  z\right)  $ and $\widetilde{\epsilon_{\varkappa}}%
:z\mapsto\epsilon_{\varkappa}\left(  -z\right)  $ of $z\in\mathbb{R}$
respectively, with $\widetilde{\epsilon_{\varkappa}}^{+}\rho_{0}%
\widetilde{\epsilon_{\varkappa}}=\rho=\epsilon_{\varkappa}^{+}\rho_{0}%
\epsilon_{\varkappa}$ due to $\rho\left(  z\right)  =\rho\left(  -z\right)  $.
Both these operators are isometries of the Hilbert space $L_{\mathfrak{h}}%
^{2}\left(  \mathbb{R},\rho\right)  $ with the scalar product
\[
\left\Vert \varphi\right\Vert _{\rho}^{2}=\int_{-\infty}^{\infty}\left\Vert
\sqrt{\rho\left(  z\right)  }\varphi\left(  z\right)  \right\Vert
^{2}\mathrm{d}z=\int_{-\infty}^{\infty}\left\Vert \sqrt{\rho\left(  z\right)
}\tilde{\varphi}\left(  z\right)  \right\Vert ^{2}\mathrm{d}z=\left\Vert
\tilde{\varphi}\right\Vert _{\rho}^{2}.
\]
into the space $\mathfrak{h}_{0}\otimes L^{2}\left(  \mathbb{R}\right)  $ of
square-integrable functions $\varphi_{0}$ with values in the Hilbert space
$\mathfrak{h}_{0}\simeq\sqrt{\rho_{0}}\mathfrak{h}$, the completion of
$\mathfrak{h}$ with respect to the norm $\left\Vert \eta\right\Vert
_{0}=\left\Vert \sqrt{\rho_{0}}\eta\right\Vert $. They are unitary if
$\rho\left(  z\right)  $ is invertible for all $z\in\mathbb{R}$ such that
$\hat{\epsilon}_{\varkappa}^{-1}=\hat{\epsilon}_{\varkappa}^{\ast}$,
$\check{\epsilon}_{\varkappa}^{-1}=\check{\epsilon}_{\varkappa}^{\ast}$,
where
\[
\epsilon_{\varkappa}^{\ast}\left(  z\right)  =\rho\left(  z\right)
^{-1}\epsilon_{\varkappa}\left(  z\right)  ^{+}\rho_{0}\text{.}%
\]
If $\hbar\hat{\gamma}_{0}=\rho_{0}^{-1/2}h\rho_{0}^{1/2}$ is an operator in
$\mathfrak{h}_{0}\otimes L^{2}\left(  \mathbb{R}\right)  $ which is equivalent
to a pseudo-differential operator $\hat{h}=h\left(  z,\frac{\hbar}{\mathrm{i}%
}\partial_{z}\right)  $ in $L_{\mathfrak{h}}^{2}\left(  \mathbb{R}\right)
=\mathfrak{h}\otimes L^{2}\left(  \mathbb{R}\right)  $, the (generalized)
function
\begin{equation}
\gamma_{\varkappa}\left(  z,\kappa\right)  =\epsilon_{\varkappa+\kappa}^{\ast
}\left(  z\right)  \gamma_{0}\left(  z,\mathrm{i}\partial_{z}\right)
\epsilon_{\varkappa+\kappa}\left(  z\right)  \equiv\gamma_{\varkappa+\kappa
}\left(  z\right)  ,\label{2.6}%
\end{equation}
is the symbol of the operator
\[
\hat{\gamma}_{\varkappa}=\hat{\epsilon}_{\varkappa}^{\ast}\hat{\gamma}_{0}%
\hat{\epsilon}_{\varkappa}\equiv\gamma_{\varkappa}\left(  z,\mathrm{i}%
\partial_{z}\right)
\]
in $L_{\mathfrak{h}}^{2}\left(  \mathbb{R},\rho\right)  $. It is defined on
the exponential functions $\epsilon_{\kappa}\left(  z\right)  =e^{-\mathrm{i}%
\kappa z}$ as the pseudo-differential operator
\[
\left[  \hat{\gamma}_{\varkappa}\epsilon_{\kappa}\eta\right]  \left(
z\right)  =\gamma_{\varkappa}\left(  z,\mathrm{i}\partial_{z}\right)
e^{-\mathrm{i}\kappa z}\eta=e^{-\mathrm{i}\kappa z}\gamma_{\varkappa}\left(
z,\kappa\right)  \eta,\;\eta\in\mathfrak{h}\text{.}%
\]

\begin{proposition}
Let $\mathcal{E}_{0}^{-}\simeq\sqrt{\rho_{0}}\mathcal{E}^{-}$ be the Hardy
class $\mathcal{E}^{-}$ of $\mathfrak{h}_{0}\otimes L^{2}\left(
\mathbb{R}\right)  $, $\mathcal{E}_{\varkappa}^{-}\subset L_{\mathfrak{h}}%
^{2}\left(  \mathbb{R},\rho\right)  $ be the Hilbert space of functions
$\varphi=\hat{\epsilon}_{\varkappa}^{\ast}\varphi_{0}$ with $\varphi_{0}%
\in\mathcal{E}_{0}^{-}$, and $\mathcal{E}_{\varkappa}^{+}=\check{\epsilon
}_{\varkappa}^{\ast}\mathcal{E}_{0}^{+}$, where $\mathcal{E}_{0}^{+}%
\simeq\sqrt{\rho_{0}}\mathcal{E}^{+}$. Let the initial boundary-value
Schr\"{o}dinger problem
\begin{align}
\mathrm{i}\partial_{t}\varphi^{t}\left(  z\right)   &  =\varepsilon
_{\varkappa}\left(  z,\mathrm{i}\partial_{z}\right)  \varphi^{t}\left(
z\right)  ,\quad\varphi^{0}\in\mathcal{E}_{\varkappa}^{-},z>0,\label{2.7}\\
\mathrm{i}\partial_{t}\tilde{\varphi}^{t}\left(  z\right)   &  =\tilde
{\varepsilon}_{\varkappa}\left(  z,\mathrm{i}\partial_{z}\right)
\tilde{\varphi}^{t}\left(  z\right)  ,\quad\tilde{\varphi}^{t}\left(
0\right)  =\sigma_{0}\varphi^{t}\left(  0\right)  ,\nonumber
\end{align}
be defined by the generators $\hat{\varepsilon}_{\varkappa},\check
{\varepsilon}_{\varkappa}$ given by the symbols $\varepsilon_{\varkappa
}\left(  z,\kappa\right)  =\varepsilon_{\varkappa+\kappa}\left(  z\right)  $,
$\tilde{\varepsilon}_{\varkappa}\left(  z,\kappa\right)  =\varepsilon
_{\varkappa-\kappa}\left(  -z\right)  $ respectively with
\[
\varepsilon_{\varkappa}\left(  z\right)  =\epsilon_{\varkappa}^{\ast}\left(
z\right)  \varepsilon_{0}\left(  \mathrm{i}\partial_{z}\right)  \epsilon
_{\varkappa}\left(  z\right)  ,
\]
where $\varepsilon_{0}\left(  \kappa\right)  =\rho_{0}^{-1/2}\varepsilon
\left(  \kappa\right)  \rho_{0}^{1/2}$ is the symmetric function of $\kappa
\in\mathbb{R}$, corresponding to the kinetic energy $e\left(  p\right)  >0$,
and $\sigma_{0}=\rho_{0}^{-1/2}\sigma\rho_{0}^{1/2}$. Then it is self-adjoint
if the initial output waves $\tilde{\varphi}^{0}$ are defined in
$\mathcal{E}_{\varkappa}^{+}$ by $\tilde{\varphi}^{0}\left(  -z\right)
=\sigma_{\varkappa}\left(  z\right)  \varphi^{0}\left(  z\right)  $, $z<0$,
where $\sigma_{\varkappa}=\epsilon_{\varkappa}^{\ast}\sigma_{0}\epsilon
_{\varkappa}$, by analytical continuation of each $\varphi_{0}^{0}%
=\hat{\epsilon}_{\varkappa}\varphi^{0}$ into the domain $\mathbb{R}^{-}$. The
solutions to (\ref{2.7}) can be written as
\begin{equation}
\varphi^{t}\left(  z\right)  =\phi^{t}\left(  z\right)  ,z\geq0,\;\tilde
{\varphi}^{t}\left(  -z\right)  =\phi^{t}\left(  z_{-}\right)  ,z\leq
0\label{2.8}%
\end{equation}
where $\phi^{t}=e^{-\mathrm{i}t\hat{\varepsilon}_{\varkappa}}\phi_{t}$,
$\phi_{t}=\varphi^{0}+\left(  \hat{\sigma}_{\varkappa}-1\right)  \hat{\pi
}_{\varkappa}^{t}\varphi^{0}$, $\hat{\sigma}_{\varkappa}$ is pointwise
multiplication by $\sigma_{\varkappa}\left(  z\right)  $, and
\[
\hat{\pi}_{\varkappa}^{t}=e^{\mathrm{i}t\hat{\varepsilon}_{\varkappa}}\hat
{1}_{0}e^{-\mathrm{i}t\hat{\varepsilon}_{\varkappa}}\equiv\pi_{\varkappa}%
^{t}\left(  z,\mathrm{i}\partial_{z}\right)
\]
is given by the symbol $\pi^{t}\left(  z,\kappa\right)  $ of the
orthoprojector $\hat{\pi}^{t}=e^{\mathrm{i}t\hat{\varepsilon}}\hat{1}%
_{0}e^{-\mathrm{i}t\hat{\varepsilon}}$ as in (\ref{2.6}).
\end{proposition}

\begin{proof}
Separating the variable $t\in\mathbb{R}$ by $\varphi^{t}=e^{-\mathrm{i}%
\varepsilon_{k}t}\varphi_{k}$, $\tilde{\varphi}^{t}=e^{-\mathrm{i}%
\varepsilon_{k}t}\tilde{\varphi}_{k}$, let us consider the stationary
Schr\"{o}dinger problem
\begin{equation}
\varepsilon_{\varkappa}\left(  z,\mathrm{i}\partial_{z}\right)  \varphi
_{k}\left(  z\right)  =\varepsilon_{k}\varphi_{k}\left(  z\right)
,\quad\tilde{\varphi}_{k}\left(  -z\right)  =\sigma_{\varkappa}\left(
z\right)  \varphi_{k}\left(  z\right)  \label{2.9}%
\end{equation}
corresponding to the given initial and boundary conditions in (\ref{2.7}).
Here $\varphi_{k}$ is extended to the domain $\mathbb{R}^{-}$ through the
analytical continuation of $\epsilon_{\varkappa}^{\ast}\varphi_{k}$ in
$\operatorname{Im}z<0$, which are the generalized eigen-functions (\ref{2.2})
of $\hat{\varepsilon}=\varepsilon\left(  \mathrm{i}\partial_{z}\right)  $ in
$\mathcal{E}_{0}^{-}$ iff $k<0$. Due to the self-adjointness of $\hat
{\varepsilon}$ in $\mathcal{E}^{-}$, the eigenfunctions $\varphi_{k}%
=\epsilon_{\varkappa+k}^{\ast}\eta_{k}$ of $\hat{\varepsilon}_{\varkappa}$ for
(\ref{2.9}) with negative $k$ form an orthocomplete set for the Hilbert space
$\mathcal{E}_{\varkappa}^{-}$, and the output eigen-functions $\tilde{\varphi
}_{k}\left(  z\right)  =\widetilde{\epsilon_{\varkappa+k}^{\ast}}\left(
z\right)  \tilde{\eta}_{k}$, where $\tilde{\eta}_{k}=\sigma_{0}\eta_{k}$ with
$\sigma_{0}=\rho_{0}^{-1/2}\sigma\rho_{0}^{1/2}$, form an orthocomplete set
for the Hilbert space $\mathcal{E}_{\varkappa}^{+}$. The solutions to
(\ref{2.7}) can be written in the form (\ref{2.3}) as
\begin{align*}
\varphi^{t}\left(  z\right)   &  =\frac{1}{2\pi}\int_{-\infty}^{0}%
e^{-\mathrm{i}\varepsilon\left(  k\right)  t}\epsilon_{\varkappa+k}^{\ast
}\left(  z\right)  f_{0}\left(  k\right)  \mathrm{d}k=\left[  e^{-\mathrm{i}%
t\hat{\varepsilon}_{\varkappa}}\varphi^{0}\right]  \left(  z\right)  ,\\
\widetilde{\varphi}^{t}\left(  z\right)   &  =\frac{1}{2\pi}\int_{-\infty}%
^{0}e^{-\mathrm{i}\varepsilon\left(  k\right)  t}\widetilde{\epsilon
_{\varkappa+k}^{\ast}}\left(  z\right)  \tilde{f}_{0}\left(  k\right)
\mathrm{d}k=\left[  e^{-\mathrm{i}t\check{\varepsilon}_{\varkappa}}%
\widetilde{\varphi}^{0}\right]  \left(  z\right)  ,
\end{align*}
where $\tilde{f}_{0}$ $\left(  k\right)  =\sigma_{0}f_{0}\left(  k\right)  $
are defined as the Fourier transforms
\[
f_{0}\left(  k\right)  =\int_{-\infty}^{\infty}\epsilon_{\varkappa+k}\left(
z\right)  \varphi^{0}\left(  z\right)  \mathrm{d}z,\;\tilde{f}_{0}\left(
k\right)  =\int_{-\infty}^{\infty}\widetilde{\epsilon_{\varkappa+k}}\left(
z\right)  \tilde{\varphi}^{0}\left(  z\right)  \mathrm{d}z,\;
\]
by the initial conditions, analytically extended on the whole line
$\mathbb{R}.$ Due to the commutativity of $\sigma$ and $\hat{\varepsilon}$
they satisfy the connection $\tilde{\varphi}^{t}\left(  -z\right)
=\sigma_{\varkappa}\left(  z\right)  \varphi^{t}\left(  z\right)  $ for all
$t$, not only for $t=0$. The time invariance of this connection and the
unitarity of the time transformation group in the Hilbert space $\mathcal{E}%
_{\varkappa}^{-}\oplus\mathcal{E}_{\varkappa}^{+}$, which follows from the
unitarity of (\ref{2.3}) in $\mathcal{E}^{\mp}\subset L_{\mathfrak{h}}%
^{2}\left(  \mathbb{R}\right)  $, means the self-adjointness of the problem
(\ref{2.7}) for the pairs $\varphi^{\mp}\in L_{\mathfrak{h}}^{2}\left(
\mathbb{R}\mathbf{,\rho}\right)  $ in the domain of the generator
$\hat{\varepsilon}_{\varkappa}\oplus\check{\varepsilon}_{\varkappa}$ with the
connection $\varphi^{+}\left(  -z\right)  =\sigma_{\varkappa}\left(  z\right)
\varphi^{-}\left(  z\right)  $. Introducing
\[
\phi^{t}\left(  z\right)  =\varphi^{t}\left(  z\right)  +1_{0}\left(
z\right)  \left(  \sigma_{\varkappa}\left(  z\right)  -1\right)  \varphi
^{t}\left(  z\right)  =\sigma_{\varkappa}\left(  z\right)  ^{1_{0}\left(
z\right)  }\varphi^{t}\left(  z\right)
\]
as in (\ref{2.4}), and taking into account that
\[
\phi^{t}\left(  z_{-}\right)  =\sigma_{\varkappa}\left(  z\right)
^{1_{0}\left(  z_{-}\right)  }\varphi^{t}\left(  z\right)  =\sigma_{\varkappa
}\left(  z\right)  ^{1-1_{0}\left(  -z\right)  }\varphi^{t}\left(  z\right)
=\sigma_{\varkappa}\left(  z\right)  ^{-1_{0}\left(  -z\right)  }%
\tilde{\varphi}^{t}\left(  -z\right)  ,
\]
we obtain the representation (\ref{2.8}) as $\varphi^{t}\left(  z\right)  $
coincides with $\phi^{t}\left(  z\right)  $ at $z\geq0$ and $\tilde{\varphi
}^{t}\left(  -z\right)  $ with $\tilde{\phi}^{t}\left(  -z\right)  =\phi
^{t}\left(  z_{-}\right)  $ at $z\leq0$.
\end{proof}

\begin{remark}
The Schr\"{o}dinger boundary value problem (\ref{2.7}) is physical in all
three aspects. First, the equation (\ref{2.7}) is invariant under the
reversion of time arrow, i.e. under the reflection $t\mapsto-t$ and an
isometric complex conjugation $\varphi\mapsto\bar{\varphi}$ together with the
input-output exchange $\varphi\leftrightarrows\tilde{\varphi}$ if $\bar
{\sigma}=\sigma^{-1}$, $\bar{\rho}_{0}=\rho_{0}$ and $\bar{\varkappa}%
=\tilde{\varkappa}$, where $\tilde{\varkappa}\left(  z\right)  =\varkappa
\left(  -z\right)  $. Second, the wave functions $\varphi^{t},\tilde{\varphi
}^{t}$ have continuous propagation in both directions of the momentum along
$z$, and at the boundary $z=0$ the momentum changes its direction but not the
magnitude (conservation of momentum) as the result of the boundary condition
$\varphi\left(  0\right)  \mapsto\tilde{\varphi}\left(  0\right)  $. And
third, the kinetic energy operator $\hat{\varepsilon}_{\varkappa}\oplus
\check{\varepsilon}_{\varkappa}$ is bounded from below as the result of
unitary transformation of $\hat{\varepsilon}\simeq\check{\varepsilon}$.
(\ref{1.3}).
\end{remark}

Indeed, from $\bar{\rho}_{0}=\rho_{0}$ and $\bar{\varepsilon}\left(
\kappa\right)  =\varepsilon\left(  \kappa\right)  =\tilde{\varepsilon}\left(
\kappa\right)  $ it follows that the symbol $\bar{\varepsilon}_{\varkappa
}\left(  z,\kappa\right)  =\overline{\varepsilon_{\varkappa-\kappa}}\left(
z\right)  $ of the complex conjugated operator $\overline{\varepsilon
_{\varkappa}}$ is given by
\[
\overline{\varepsilon_{\varkappa}}\left(  z\right)  =\epsilon_{-\bar
{\varkappa}}^{\ast}\left(  z\right)  \varepsilon_{0}\left(  \mathrm{i}%
\partial_{z}\right)  \epsilon_{-\bar{\varkappa}}\left(  z\right)
=\epsilon_{-\tilde{\varkappa}}^{\ast}\left(  z\right)  \varepsilon_{0}\left(
\mathrm{i}\partial_{z}\right)  \epsilon_{-\tilde{\varkappa}}\left(  z\right)
=\widetilde{\varepsilon_{\varkappa}}\left(  z\right)
\]
if $\bar{\varkappa}=\tilde{\varkappa}$, as $\overline{\epsilon_{\varkappa}%
}\left(  z\right)  =\epsilon_{-\bar{\varkappa}}\left(  z\right)  $ and
$\epsilon_{-\tilde{\varkappa}}\left(  z\right)  =\widetilde{\epsilon
_{\varkappa}}\left(  z\right)  $ in (\ref{2.5}). Therefore $\bar{\varepsilon
}_{\varkappa}\left(  z,\kappa\right)  =$ $\tilde{\varepsilon}_{\varkappa
}\left(  z,\kappa\right)  $, where $\tilde{\varepsilon}_{\varkappa}\left(
z,\kappa\right)  =\widetilde{\varepsilon_{\varkappa-\kappa}}\left(  z\right)
$ is the symbol for the kinetic energy operator $\check{\varepsilon
}_{\varkappa}=\widetilde{\hat{\varepsilon}_{\varkappa}}$ for the output wave
$\tilde{\varphi}$. Thus the time reversion with complex conjugation in
(\ref{2.7}) is equivalent to the input-output interchange $\left(  \varphi
^{t},\tilde{\varphi}^{t}\right)  \mapsto\left(  \tilde{\varphi}^{t}%
,\varphi^{t}\right)  $ which preserves the connection between $\varphi^{t}$
and $\tilde{\varphi}^{t}$ as
\[
\overline{\sigma_{\varkappa}}\left(  z\right)  =\epsilon_{-\bar{\varkappa}%
}^{\ast}\left(  z\right)  \overline{\sigma_{0}}\epsilon_{-\bar{\varkappa}%
}\left(  z\right)  =\epsilon_{-\tilde{\varkappa}}^{\ast}\left(  z\right)
\sigma_{0}^{-1}\epsilon_{-\tilde{\varkappa}}\left(  z\right)  =\widetilde
{\sigma_{\varkappa}}\left(  z\right)  ^{-1},
\]
where $\widetilde{\sigma_{\varkappa}}\left(  z\right)  =\sigma_{\varkappa
}\left(  -z\right)  $ due to $\overline{\sigma_{0}}=\sigma_{0}^{-1}$.

Note that the orthoprojectors $\hat{\pi}^{t}$ commute with $\sigma_{0}$ on
$\mathcal{E}_{0}^{-}$ applied pointwisely as $\left[  \sigma\phi\right]
\left(  z\right)  =\sigma\phi\left(  z\right)  $ such that
\[
\sigma^{\hat{\pi}^{t}}=I+\left(  \sigma-1\right)  \hat{\pi}^{t},\quad
\sigma^{-\hat{1}_{0}}=\hat{1}+\hat{1}_{0}\left(  \sigma^{-1}-1\right)
\]
are well defined as unitary operators on $\mathfrak{h}_{0}\otimes L^{2}\left(
\mathbb{R}\right)  $. However in general they do not commute with each other
and are not orthogonal to $\hat{1}_{0}^{\bot}=I-\hat{1}_{0}$, and thus the
unitary cocycle $\upsilon_{t}=\sigma^{\hat{\pi}^{t}}\sigma^{-\hat{1}_{0}}$
does not coincide with $\sigma^{\hat{1}_{t}-\hat{1}_{0}}$. Indeed, $\hat{\pi
}^{t}=\hat{1}_{0}^{t}$ can be represented as
\begin{equation}
\hat{\pi}^{t}=\int_{-\infty}^{0}\hat{\delta}_{r}^{t}\mathrm{d}r,\quad
\hat{\delta}_{r}^{t}=\frac{1}{2\pi}\int_{-\infty}^{\infty}e^{\mathrm{i}\kappa
r}\hat{\epsilon}_{\kappa}^{t}\mathrm{d}\kappa\label{2.10}%
\end{equation}
where $\hat{\epsilon}_{\kappa}^{t}=e^{\mathrm{i}t\hat{\varepsilon}}%
\hat{\epsilon}_{\kappa}e^{-\mathrm{i}t\hat{\varepsilon}}$ is the Heisenberg
transformation of the unitary multiplication group $\hat{\epsilon}_{\kappa
},\kappa\in\mathbb{R}$ by the exponential function $\epsilon_{\kappa}:z\mapsto
e^{-\mathrm{i}\kappa z}$. The unitary operators $\hat{\epsilon}_{\kappa}^{t}$
are defined on the harmonic eigen-functions of $\hat{\varepsilon}$ as shifts
\begin{align*}
\left[  \hat{\epsilon}_{\kappa}^{t}\varphi_{k}\right]  \left(  z\right)   &
=e^{\mathrm{i}t\varepsilon\left(  \mathrm{i}\partial_{z}\right)
}e^{-\mathrm{i}\kappa z}e^{-\mathrm{i}t\varepsilon\left(  k\right)  t}%
\varphi_{k}\left(  z\right)  \\
&  =e^{\mathrm{i}t\left(  \varepsilon\left(  k-\kappa\right)  -\varepsilon
\left(  k\right)  \right)  }e^{-\mathrm{i}\kappa z}\varphi_{k}\left(
z\right)  ,
\end{align*}
of the negative $k$ of $\varphi_{k}$ to $k-\kappa$ of $\epsilon_{\kappa
}\varphi_{k}$, $\kappa\in\mathbb{R}$. The commutativity of $\hat{1}_{0}^{t}$
and $\hat{1}_{0}$ takes place when $\hat{\epsilon}_{\kappa}^{t}$\ commutes
with $\hat{\epsilon}_{\kappa^{\circ}}$, $\kappa^{\circ}\in\mathbb{R}$, i.e.
when
\[
\varepsilon\left(  k-\kappa^{\circ}-\kappa\right)  -\varepsilon\left(
k-\kappa^{\circ}\right)  =\varepsilon\left(  k-\kappa\right)  -\varepsilon
\left(  k\right)  ,\quad\forall\kappa^{\circ}\text{.}%
\]
This would be possible if $\varepsilon\left(  k\right)  $\ were an affine
function, say $\varepsilon\left(  k\right)  =\varepsilon_{0}+k$. In this case
$\hat{\epsilon}_{\kappa}^{t}$ is the multiplication by $\epsilon_{\kappa
}\left(  z-t\right)  $, $\hat{\pi}^{t}$ is the multiplication $\hat{1}_{t}$ by
$1_{t}\left(  z\right)  =1_{0}\left(  z-t\right)  $. This is the case when the
correspondent cocycle
\[
\upsilon_{t}=\sigma^{\hat{1}_{t}-\hat{1}_{0}}=I+\left(  \sigma-1\right)
\left(  \hat{1}_{t}-\hat{1}_{0}\right)
\]
coincides with the unitary evolution in the interaction picture
\[
e^{\mathrm{i}\hat{\varepsilon}t}e^{\left(  \hat{\delta}\ln\sigma
-\mathrm{i}\hat{\varepsilon}\right)  t}=\overleftarrow{\exp}\left[  \int
_{0}^{t}\hat{\delta}^{r}\mathrm{d}r\ln\sigma\right]  ,\quad\hat{\delta}%
^{r}=\frac{1}{2\pi}\int_{-\infty}^{\infty}\hat{\epsilon}_{\kappa}%
^{r}\mathrm{d}\kappa
\]
for the $\delta$-function interaction potential $\Phi\left(  z\right)
=\mathrm{i}\hbar\delta\left(  z\right)  \ln\sigma$. Indeed, if $\hat
{\varepsilon}=\varepsilon+\mathrm{i}\partial_{z}$, $\hat{\delta}^{r}$ is the
multiplication $\hat{\delta}_{r}$ by $\delta_{r}\left(  z\right)
=\delta\left(  z-r\right)  $, and $\int_{0}^{t}\hat{\delta}^{r}\mathrm{d}r$ is
the multiplication operator by $\Delta_{0}^{t}=\int_{0}^{t}\delta
_{r}\mathrm{d}r=1_{t}-1_{0}$. However the affine form of $\varepsilon\left(
k\right)  $ for all $k\in\mathbb{R}$ contradicts to the physical assumption of
positivity and symmetricity $\varepsilon\left(  \pm k\right)  =\varepsilon
\left(  k\right)  >0$ of the reflection boundary value problem.

Thus the Hamiltonian boundary value problem in $\mathbb{R}^{+}$, corresponding
to the free propagation of input and output waves in the opposite directions
with the unitary reflection at $z=0$, in general cannot be reduced to the
propagation problem in $\mathbb{R}$ with the unitary transition from $z=0$ to
$z=0_{-}$ corresponding to a $\delta$-function potential on $\mathbb{R}$.
However we shall see now that at the ultra relativistic limit this boundary
value problem is equivalent to the $\delta$-potential problem for the toy model.

\section{The ultra relativistic limit.}

We shall assume here that the symmetric positive kinetic energy $e\left(
p\right)  $ has the relativistic form $\left\vert p\right\vert $, or more
generally, $e\left(  p\right)  =\sqrt{p^{2}+\hbar^{2}\mu^{2}}$ as it was
suggested in (\ref{2.1}). It corresponds to the finite bounds $v_{\mp}=\mp1$
of the velocity $v\left(  p\right)  =\varepsilon^{\prime}\left(  p\right)  $
at $p\longrightarrow\mp\infty$. Note that the phase speed
\[
\varsigma_{\kappa}=\varepsilon\left(  \kappa\right)  /\kappa=\sqrt
{1+\varepsilon^{2}/\kappa^{2}}=\left\vert v\left(  \hbar\kappa\right)
^{-1}\right\vert ,
\]
for the momenta $p=\mp\hbar\kappa,\kappa>0$ of the harmonic eigen-waves
\[
e^{-\mathrm{i}\varepsilon_{\kappa}t}\epsilon_{\kappa}\left(  z\right)
=e^{-\mathrm{i}\kappa\left(  \varsigma_{\kappa}t+z\right)  },\quad
e^{-\mathrm{i}\varepsilon_{\kappa}t}\tilde{\epsilon}_{\kappa}\left(  z\right)
=e^{-\mathrm{i}\kappa\left(  \varsigma_{\kappa}t-z\right)  }%
\]
has also the limit $\varsigma=1$ at $\kappa\longrightarrow\infty$. Therefore
one should expect that the rapidly oscillating input and output waves
\begin{equation}
\varphi^{t}\left(  z\right)  =e^{-\mathrm{i}\kappa\left(  t+z\right)  }%
\psi^{t}\left(  z\right)  ,\quad\tilde{\varphi}^{t}\left(  z\right)
=e^{-\mathrm{i}\kappa\left(  t-z\right)  }\tilde{\psi}^{t}\left(  z\right)
,\label{3.1}%
\end{equation}
in the ultra relativistic limit $p\longrightarrow\mp\infty$ will propagate as
the plane waves with
\begin{equation}
\psi^{t}\left(  z\right)  =\psi\left(  z+t\right)  \equiv e^{t\partial_{z}%
}\psi,\quad\tilde{\psi}^{t}\left(  z\right)  =\tilde{\psi}\left(  z-t\right)
\equiv e^{t\tilde{\partial}_{z}}\tilde{\psi}\label{3.5}%
\end{equation}
if the initial conditions are prepared in this form with slowly changing
amplitudes $\psi,\tilde{\psi}\in L_{\mathfrak{h}}^{2}\left(  \mathbb{R}%
\right)  $. This propagation will reproduce the boundary-reflection dynamics
$\tilde{\psi}^{t}\left(  0\right)  =\sigma\psi\left(  0\right)  $ on the half
line $\mathbb{R}^{+}\ni z=s$ if the initial wave amplitudes are connected by
$\tilde{\psi}\left(  -z\right)  =\sigma\psi\left(  z\right)  $ for all
$z\in\mathbb{R}$. In particular, the solutions $\psi^{t}\left(  s\right)
=\psi\left(  s+t\right)  $, $\tilde{\psi}^{t}\left(  s\right)  =0,t<s$,
\[
\tilde{\psi}^{t}\left(  s\right)  =\tilde{\psi}^{t-s}\left(  0\right)
=\sigma\psi^{t-s}\left(  0\right)  =\sigma\psi\left(  t-s\right)  ,t>s
\]
to this Hamiltonian boundary value problem with the input wave functions
\[
\psi\left(  z\right)  =\sqrt{\rho\left(  z\right)  }\eta,z>0,\quad\psi\left(
z\right)  =0,z\leq0,
\]
for the initial state-vectors $\eta\in\mathfrak{h}$ will correspond to the
single-jump stochastic dynamics in the positive direction of $t$ with respect
to the probability density $\rho>0,$ $\int_{0}^{\infty}\rho\left(  s\right)
\mathrm{d}s=1$.

Below we give a precise formulation and proof of this conjecture in a more
general framework which is needed for the derivation of quantum stochastic
evolution as the boundary value problem in second quantization. But first let
us introduce the notations and illustrate this limit in this simple case.

In the following we shall use the notion of the inductive limit of an
increasing family $\left(  \mathcal{E}_{\kappa}\right)  _{\kappa>0}$ of
Hilbert subspaces $\mathcal{E}_{\kappa}\subseteq\mathcal{E}_{\kappa^{\prime}%
},\kappa<\kappa^{\prime}$. It is defined as the union $\mathcal{E}%
=\cup\mathcal{E}_{\kappa}$ equipped with the inductive convergence which is
the uniform convergence in one of the subspaces $\mathcal{E}_{\kappa}$. A
sequence $\left(  \psi_{\nu}\right)  $ converges in $\mathcal{E}$ if there
exists a $\kappa$ that $\psi_{\nu}\in\mathcal{E}_{\kappa}$ for all $\nu
>\nu_{\kappa}$, and $\left(  \psi_{\nu}\right)  _{\nu>\nu_{\kappa}}$ converges
in $\mathcal{E}_{\kappa}$. The inductive convergence is stronger than the
convergence in the uniform completion $\mathcal{K}=\overline{\mathcal{E}}$.
Therefore the dual convergence is weaker then the convergence in $\mathcal{K}%
$. The inductive operator convergence in $\mathcal{E}$ is defined as the
operator convergence on each $\mathcal{E}_{\kappa}$ into one of $\mathcal{E}%
_{\kappa^{\prime}}\subseteq\mathcal{K}$.

Let $\mathcal{G}^{-}=\cup\mathcal{E}_{\kappa}^{-}$, $\mathcal{G}^{+}%
=\cup\mathcal{E}_{\kappa}^{+}$ be the inductive limits for the increasing
family $\left(  \mathcal{E}_{\kappa}^{-},\mathcal{E}_{\kappa}^{+}\right)
_{\kappa>0}$ of Hardy classes $\mathcal{E}_{\kappa}^{-}=\hat{\epsilon}%
_{\kappa}^{\ast}\mathcal{E}^{-}\supset\mathcal{E}_{\kappa^{\circ}}^{-}$,
$\mathcal{E}_{\kappa}^{+}=\check{\epsilon}_{\kappa}^{\ast}\mathcal{E}%
^{+}\supset\mathcal{E}_{\kappa^{\circ}}^{+}$, $\kappa^{\circ}<\kappa$ in the
notations of the previous section. Both $\mathcal{G}^{-},\mathcal{G}^{+}$ are
dense in $L_{\mathfrak{h}}^{2}\left(  \mathbb{R}\right)  $, consist of the
square-intergable $\mathfrak{h}$-valued functions $\psi\in\mathcal{G}^{-}$,
$\tilde{\psi}\in\mathcal{G}^{+}$ having zero Fourier amplitudes
\[
g\left(  k\right)  =\int_{-\infty}^{\infty}e^{-\mathrm{i}kz}\psi\left(
z\right)  \mathrm{d}z,\quad\tilde{g}\left(  k\right)  =\int_{-\infty}^{\infty
}e^{\mathrm{i}kz}\tilde{\psi}\left(  z\right)  \mathrm{d}z
\]
for all $k\geq\kappa$ with sufficiently large $\kappa>0$. If $\psi
\in\mathcal{E}_{\kappa}^{-}$ and $\tilde{\psi}\in\mathcal{E}_{\kappa}^{+}$,
then $\varphi=\epsilon_{\kappa}\psi\in\mathcal{E}^{-}$, $\tilde{\varphi
}=\tilde{\epsilon}_{\kappa}\tilde{\psi}\in\mathcal{E}^{+}$, and the free
propagation (\ref{2.3}) can be written in the form (\ref{3.1}) with
\begin{align*}
\psi^{t} &  =e^{\mathrm{i}\kappa t}\hat{\epsilon}_{\kappa}^{\ast}\varphi
^{t}=\hat{\epsilon}_{\kappa}^{\ast}e^{-\mathrm{i}\left(  \hat{\varepsilon
}-\kappa1\right)  t}\hat{\epsilon}_{\kappa}\psi\equiv\psi_{\kappa}^{t},\\
\tilde{\psi}^{t} &  =e^{\mathrm{i}\kappa t}\check{\epsilon}_{\kappa}^{\ast
}\tilde{\varphi}^{t}=\check{\epsilon}_{\kappa}^{\ast}e^{-\mathrm{i}\left(
\check{\varepsilon}-\kappa1\right)  t}\check{\epsilon}_{\kappa}\tilde{\psi
}\equiv\tilde{\psi}_{\kappa}^{t}.
\end{align*}
These unitary transformations in $\mathcal{E}_{\kappa}^{-}$ and in
$\mathcal{E}_{\kappa}^{+}$, written as
\begin{equation}
\psi_{\kappa}^{t}\left(  z\right)  =e^{-\mathrm{i}t\omega_{\kappa}\left(
\mathrm{i}\partial_{z}\right)  }\psi\left(  z\right)  ,\quad\tilde{\psi
}_{\kappa}^{t}\left(  z\right)  =e^{-\mathrm{i}t\tilde{\omega}_{\kappa}\left(
\mathrm{i}\partial_{z}\right)  }\tilde{\psi}\left(  z\right)  ,\label{3.2}%
\end{equation}
are generated by the self-adjoint operators
\begin{align}
\omega_{\kappa}\left(  \mathrm{i}\partial_{z}\right)   &  =e^{\mathrm{i}\kappa
z}\left(  \varepsilon\left(  \mathrm{i}\partial_{z}\right)  -\kappa\right)
e^{-\mathrm{i}\kappa z}=\varepsilon\left(  \kappa+\mathrm{i}\partial
_{z}\right)  -\kappa,\;\label{3.4}\\
\tilde{\omega}_{\kappa}\left(  \mathrm{i}\partial_{z}\right)   &
=e^{-\mathrm{i}\kappa z}\left(  \varepsilon\left(  \mathrm{i}\partial
_{z}\right)  -\kappa\right)  e^{\mathrm{i}\kappa z}=\varepsilon\left(
\kappa-\mathrm{i}\partial_{z}\right)  -\kappa\nonumber
\end{align}
They leave all subspaces $\mathcal{E}_{\kappa^{\circ}}^{-}$ and $\mathcal{E}%
_{\kappa^{\circ}}^{+}$ invariant respectively, however their generators
$\hat{\omega}_{\kappa},\check{\omega}_{\kappa}$ are not positive definite for
a sufficiently large $\kappa$, and are not unitary equivalent for different
$\kappa$ as
\begin{align*}
\hat{\epsilon}_{\varkappa}\hat{\omega}_{\kappa}\hat{\epsilon}_{\varkappa
}^{\ast} &  =\hat{\varepsilon}_{\kappa^{\circ}}-\kappa1=\hat{\omega}%
_{\kappa^{\circ}}-\varkappa1,\quad\\
\check{\epsilon}_{\varkappa}\check{\omega}_{\kappa}\check{\epsilon}%
_{\varkappa}^{\ast} &  =\check{\varepsilon}_{\kappa^{\circ}}-\kappa
1=\check{\omega}_{\kappa^{\circ}}-\varkappa1,
\end{align*}
where $\varkappa=\kappa-\kappa^{\circ}$. Thus we have to prove that the
propagation (\ref{3.3}) has the inductive limit form of plane propagation
(\ref{3.5}) at $\kappa\longrightarrow\infty$ corresponding to the Dirac form
of the limits
\[
\lim_{\kappa\rightarrow\infty}\omega_{\kappa}\left(  \mathrm{i}\partial
_{z}\right)  =\mathrm{i}\partial_{z},\quad\lim_{\kappa\rightarrow\infty}%
\tilde{\omega}_{\kappa}\left(  \mathrm{i}\partial_{z}\right)  =-\mathrm{i}%
\partial_{z}%
\]
for the Schr\"{o}dinger generators (\ref{3.3}).

Another fact which we are going to prove for obtaining the single-jump
stochastic limit is that the truncated wave
\[
\chi_{\kappa}^{t}=e^{-\mathrm{i}t\hat{\omega}_{\kappa}}\chi_{\kappa,t}%
,\quad\chi_{\kappa,t}=\psi+\left(  1-\sigma\right)  \hat{\pi}_{\kappa}^{t}\psi
\]
representing the pair (\ref{3.3}) on the half-line $\mathbb{R}^{+}\ni z$ as in
(\ref{2.8}), has the discontinuous limit
\begin{equation}
\chi^{t}\left(  z\right)  =\chi_{t}\left(  z+t\right)  ,\quad\chi_{t}%
=\psi+\left(  1-\sigma\right)  \hat{1}_{t}\psi.\label{3.6}%
\end{equation}
Here $\hat{1}_{t}=e^{-\mathrm{i}t\partial_{z}}\hat{1}_{0}e^{\mathrm{i}%
t\partial_{z}}$ is pointwise multiplication by the characteristic function
$1_{t}$ of the interval $-\infty<z<t$ which we shall obtain as the inductive
limit of the orthoprojector
\begin{equation}
\hat{\pi}_{\kappa}^{t}=e^{\mathrm{i}t\varepsilon\left(  \kappa+\mathrm{i}%
\partial_{z}\right)  }\hat{1}_{0}e^{-\mathrm{i}t\varepsilon\left(
\kappa+\mathrm{i}\partial_{z}\right)  }=e^{\mathrm{i}t\hat{\omega}_{\kappa}%
}\hat{1}_{0}e^{-\mathrm{i}t\hat{\omega}_{\kappa}}\label{3.4}%
\end{equation}
at $\kappa\longrightarrow\infty$. This results are formulated in the following
proposition in full generality and notation of the proposition 2.

\begin{proposition}
Let $\mathcal{G}_{0}^{-}\simeq\sqrt{\rho_{0}}\mathcal{G}^{-}$ be the Hilbert
inductive limit of Hardy classes $\hat{\epsilon}_{\kappa}^{\ast}%
\mathcal{E}_{0}^{-}\subset\mathfrak{h}_{0}\otimes L^{2}\left(  \mathbb{R}%
\right)  $, $\mathcal{G}_{\varkappa}^{-}\subset L_{\mathfrak{h}}^{2}\left(
\mathbb{R},\rho\right)  $ be the Hilbert space of functions $\psi
=\hat{\epsilon}_{\varkappa}^{\ast}\psi_{0}$ with $\psi_{0}\in\mathcal{G}%
_{0}^{-}$, and $\mathcal{G}_{\varkappa}^{+}=\check{\epsilon}_{\varkappa}%
^{\ast}\mathcal{G}_{0}^{+}$, where $\mathcal{G}_{0}^{+}\simeq\sqrt{\rho_{0}%
}\mathcal{G}^{+}$. Let the initial boundary-value Schr\"{o}dinger problem
\begin{align}
\mathrm{i}\partial_{t}\psi_{\kappa}^{t}\left(  z\right)   &  =\omega
_{\varkappa,\kappa}\left(  z,\mathrm{i}\partial_{z}\right)  \psi_{\kappa}%
^{t}\left(  z\right)  ,\quad\psi_{\kappa}^{0}=\psi\in\mathcal{G}_{\varkappa
}^{-},z>0,\label{3.7}\\
\mathrm{i}\partial_{t}\tilde{\psi}_{\kappa}^{t}\left(  z\right)   &
=\tilde{\omega}_{\varkappa,\kappa}\left(  z,\mathrm{i}\partial_{z}\right)
\tilde{\psi}_{\kappa}^{t}\left(  z\right)  ,z>0,\;\tilde{\psi}_{\kappa}%
^{t}\left(  0\right)  =\sigma_{0}\psi_{\kappa}^{t}\left(  0\right)  ,\nonumber
\end{align}
be defined by the generators
\[
\hat{\omega}_{\varkappa,\kappa}=\hat{\epsilon}_{\varkappa}^{\ast}\rho
_{0}^{-1/2}\hat{\omega}_{\kappa}\rho_{0}^{1/2}\hat{\epsilon}_{\varkappa}%
,\quad\check{\omega}_{\varkappa,\kappa}=\check{\epsilon}_{\varkappa}^{\ast
}\rho_{0}^{-1/2}\check{\omega}_{\kappa}\rho_{0}^{1/2}\check{\epsilon
}_{\varkappa}%
\]
with the symbols $\omega_{\kappa},\tilde{\omega}_{\kappa}$ given in
(\ref{3.4}), (\ref{2.1}), and the initial $\tilde{\psi}_{\kappa}^{0}%
=\tilde{\psi}$ defined in $\mathcal{G}_{\varkappa}^{+}$ as $\tilde{\psi
}\left(  -z\right)  =\sigma_{\varkappa}\left(  z\right)  \psi\left(  z\right)
$, $z<0$ by analytical continuation of each $\psi_{0}=\hat{\epsilon
}_{\varkappa}\psi$ into the domain $\mathbb{R}^{-}$. Then the solutions to
(\ref{3.7}) inductively converge to
\begin{equation}
\psi^{t}\left(  z\right)  =\chi^{t}\left(  z\right)  ,z\geq0,\;\tilde{\psi
}^{t}\left(  -z\right)  =\chi^{t}\left(  z_{-}\right)  ,z\leq0\label{3.8}%
\end{equation}
where $\chi^{t}\left(  z\right)  =\epsilon_{\varkappa}\left(  z,z+t\right)
\chi_{t}\left(  z+t\right)  $, and
\[
\epsilon_{\varkappa}\left(  z,z+t\right)  =\epsilon_{\varkappa}^{\ast}\left(
z\right)  \epsilon_{\varkappa}\left(  z+t\right)  ,\;\chi_{t}=\psi+\left(
\hat{\sigma}_{\varkappa}-1\right)  \hat{1}_{t}\psi.
\]

\end{proposition}

\begin{proof}
First let us note that the generators in (\ref{3.7}) have the formal limits
\begin{align*}
\lim_{\kappa\rightarrow\infty}\left[  \hat{\omega}_{\varkappa,\kappa}%
\psi\right]  \left(  z\right)   &  =\epsilon_{\varkappa}^{\ast}\left(
z\right)  \mathrm{i}\partial_{z}\left[  \epsilon_{\varkappa}{}\psi\right]
\left(  z\right)  =\left(  \varkappa\left(  z\right)  +\mathrm{i}\partial
_{z}\right)  \psi\left(  z\right)  ,\\
\lim_{\kappa\rightarrow\infty}\left[  \check{\omega}_{\varkappa,\kappa}%
\tilde{\psi}\right]  \left(  z\right)   &  =\widetilde{\epsilon_{\varkappa
}^{\ast}}\left(  z\right)  \mathrm{i}\tilde{\partial}_{z}\left[
\tilde{\epsilon}_{\varkappa}{}\tilde{\psi}\right]  \left(  z\right)  =\left(
\tilde{\varkappa}\left(  z\right)  +\mathrm{i}\tilde{\partial}_{z}\right)
\tilde{\psi}\left(  z\right)
\end{align*}
with $\tilde{\partial}_{z}=-\partial_{z}$, $\tilde{\varkappa}\left(  z\right)
=\varkappa\left(  -z\right)  $. This follows from (\ref{3.4}), and from
(\ref{2.5}) and $\partial_{z}\widetilde{\epsilon_{\varkappa}}=\mathrm{i}%
\tilde{\varkappa}\widetilde{\epsilon_{\varkappa}}$ as $\widetilde
{\epsilon_{\varkappa}}\left(  z\right)  =\epsilon_{-\tilde{\varkappa}}\left(
z\right)  $. Thus we have to prove that the solutions to (\ref{3.7}) have the
limits $\psi=\lim\psi_{\kappa}$, $\tilde{\psi}=\lim\tilde{\psi}_{\kappa}$ in
$\mathcal{G}_{\varkappa}^{\mp}$ coinciding with the solutions to the Dirac
boundary value problem
\begin{align*}
\mathrm{i}\partial_{t}\psi^{t}\left(  z\right)   &  =\left(  \varkappa\left(
z\right)  +\mathrm{i}\partial_{z}\right)  \psi^{t}\left(  z\right)  ,\quad
\psi^{0}=\psi\in\mathcal{G}_{\varkappa}^{-},z>0,\\
\mathrm{i}\partial_{t}\tilde{\psi}^{t}\left(  z\right)   &  =\left(
\tilde{\varkappa}\left(  z\right)  +\mathrm{i}\tilde{\partial}_{z}\right)
\tilde{\psi}^{t}\left(  z\right)  ,z>0,\;\tilde{\psi}^{t}\left(  0\right)
=\sigma_{0}\psi^{t}\left(  0\right)
\end{align*}
with the initial $\tilde{\psi}^{0}$ analytically defined as $\tilde{\psi}%
^{0}\left(  -z\right)  =\sigma_{\varkappa}\left(  z\right)  \psi^{0}\left(
z\right)  $ in order to keep the solution $\tilde{\psi}^{t}$ also in
$\mathcal{G}_{\varkappa}^{-}$ for all $t$.

Let us do this using the isomorphisms $\hat{\epsilon}_{\varkappa}$
$\check{\epsilon}_{\varkappa}$ of the dense subspaces $\mathcal{G}_{\varkappa
}^{\mp}\subset L_{\mathfrak{h}}^{2}\left(  \mathbb{R},\rho\right)  $ and
$\mathcal{G}_{0}^{\mp}\subset\mathfrak{h}_{0}\otimes L^{2}\left(
\mathbb{R}\right)  $. Due to this the boundary value problem (\ref{3.7}) is
equivalent to
\begin{align*}
\mathrm{i}\partial_{t}\psi_{0,\kappa}^{t}\left(  z\right)   &  =\omega
_{0,\kappa}\left(  \mathrm{i}\partial_{z}\right)  \psi_{0,\kappa}^{t}\left(
z\right)  ,\quad\psi_{0,\kappa}^{0}=\psi_{0}\in\mathcal{G}_{0}^{-},z>0\\
\mathrm{i}\partial_{t}\tilde{\psi}_{0,\kappa}^{t}\left(  z\right)   &
=\tilde{\omega}_{0,\kappa}\left(  \mathrm{i}\partial_{z}\right)  \tilde{\psi
}_{0,\kappa}^{t}\left(  z\right)  ,z>0,\;\tilde{\psi}_{0,\kappa}^{t}\left(
0\right)  =\sigma_{0}\tilde{\psi}_{0,\kappa}^{t}\left(  0\right)  ,
\end{align*}
with $\omega_{0,\kappa}\left(  -k\right)  =\varepsilon_{0}\left(
\kappa-k\right)  -\kappa=\tilde{\omega}_{0,\kappa}\left(  k\right)  $,
$\varepsilon_{0}\left(  \kappa\right)  =\rho_{0}^{-1/2}\varepsilon\left(
\kappa\right)  \rho_{0}^{1/2}$, and $\tilde{\psi}_{0,\kappa}^{0}\left(
-z\right)  =\sigma_{0,\kappa}\left(  z\right)  \psi_{0}\left(  z\right)  $
with $\sigma_{0,\kappa}=\epsilon_{\kappa}^{\ast}\sigma_{0}\epsilon_{\kappa
}=\sigma_{0}$ for any scalar $\kappa$. Thus we are to find the ultra
relativistic limit of the solutions
\begin{align}
\left[  e^{-\mathrm{i}t\hat{\omega}_{\kappa}}\psi_{0}\right]  \left(
z\right)   &  =\frac{1}{2\pi}\int_{-\infty}^{\kappa}e^{-\mathrm{i}\left(
t\omega_{\kappa}\left(  -k\right)  -kz\right)  }g_{0}\left(  k\right)
\mathrm{d}k,\label{3.9}\\
\left[  e^{-\mathrm{i}t\check{\omega}_{\kappa}}\tilde{\psi}_{0}\right]
\left(  z\right)   &  =\frac{1}{2\pi}\int_{-\infty}^{\kappa}e^{-\mathrm{i}%
\left(  t\omega_{\kappa}\left(  -k\right)  +kz\right)  }\tilde{g}_{0}\left(
k\right)  \mathrm{d}k,\nonumber
\end{align}
with $\tilde{g}_{0}$ $\left(  k\right)  =\sigma_{0}g_{0}\left(  k\right)  $ at
$\kappa\longrightarrow\infty$. Here the Fourier amplitudes
\[
g_{0}\left(  k\right)  =\int_{-\infty}^{\infty}e^{-\mathrm{i}kz}\psi
_{0}\left(  z\right)  \mathrm{d}z,\quad\tilde{g}_{0}\left(  k\right)
=\int_{-\infty}^{\infty}e^{\mathrm{i}kz}\tilde{\psi}_{0}\left(  z\right)
\mathrm{d}z,\;
\]
are defined by analytical continuation of the initial conditions $\psi_{0}%
\in\hat{\epsilon}_{\kappa^{\circ}}^{\ast}\mathcal{E}_{0}^{-}$, $\tilde{\psi
}_{0}\in\check{\epsilon}_{\kappa^{\circ}}^{\ast}\mathcal{E}_{0}^{+}$ for a
$\kappa^{\circ}<\kappa$ such that the integration in (\ref{3.9}) can be
restricted to $k<\kappa^{\circ}$ due to $g_{0}\left(  k\right)  =0=\tilde
{g}_{0}\left(  k\right)  $ for all $k\geq\kappa^{\circ}$. Therefore the proof
that the unitary evolution (\ref{3.9}) inductively converges to the plane
propagation $e^{t\partial_{z}}\psi_{0},e^{t\tilde{\partial}_{z}}\tilde{\psi
}_{0}$ resolving this problem at $\kappa\longrightarrow\infty$ can be reduced
to finding an estimate of the integral
\[
I\left(  \kappa^{\circ},\kappa\right)  =\frac{1}{2\pi}\int_{-\infty}%
^{\kappa^{\circ}}\left\Vert \left(  e^{-\mathrm{i}\left(  k+\omega_{\kappa
}\left(  -k\right)  \right)  t}-1\right)  g\left(  k\right)  \right\Vert
^{2}\mathrm{d}k.
\]
It gives the value to the mean square distances
\[
\left\Vert e^{-t\partial_{z}}\psi_{0,\kappa}^{t}-\psi_{0}\right\Vert _{0}%
^{2}=I\left(  \kappa^{\circ},\kappa\right)  =\left\Vert e^{-t\tilde{\partial
}_{z}}\tilde{\psi}_{0,\kappa}^{t}-\tilde{\psi}_{0}\right\Vert _{0}^{2}%
\]
of $\psi_{0,\kappa}^{t}\left(  z-t\right)  $ from $\psi_{0}\in\hat{\epsilon
}_{\kappa^{\circ}}^{\ast}\mathcal{E}_{0}^{-}$ having the amplitude
$g_{0}\left(  k\right)  =\rho_{0}^{-1/2}g\left(  k\right)  $ and of
$\tilde{\psi}_{0,\kappa}^{t}\left(  z+t\right)  $ from $\tilde{\psi}_{0}%
\in\check{\epsilon}_{\kappa^{\circ}}^{\ast}\mathcal{E}_{0}^{+}$ having the
amplitude $\tilde{g}_{0}\left(  k\right)  =\sigma_{0}g_{0}\left(  k\right)  $.

To this end we shall use the inequality
\[
\left(  \varkappa^{2}+\mu^{2}\right)  ^{1/2}-\varkappa<\frac{1}{2}\frac
{\mu^{2}}{\varkappa},\quad\forall\varkappa>\left|  \mu\right|
\]
for the monotonously increasing function $k+\omega_{\kappa}\left(  -k\right)
<\kappa^{\circ}+\omega_{\kappa}\left(  -\kappa^{\circ}\right)  $ of
$k<\kappa^{\circ}$. We shall treat separately the three cases in (\ref{2.1}):
the scalar massless case $\mu_{0}=0$ when $\varepsilon\left(  k\right)
=\left|  k\right|  $, the boundedness case $\left|  \mu\right|  \leq m$ when
$\varepsilon\left(  k\right)  \leq\sqrt{k^{2}+m^{2}}$ as in the scalar case
with $\mu=\mu_{0}>0$, and the general vector case when $\varepsilon\left(
k\right)  =\left(  k^{2}+\mu_{0}^{2}-\nabla^{2}\right)  ^{1/2}$.

In the first case $k+\omega_{\kappa}\left(  -k\right)  =k-\kappa+\left|
\kappa-k\right|  =0$ for all $\kappa\geq0$ and $k<\kappa$. Thus the plane wave
propagation
\[
\psi_{0,\kappa}^{t}\left(  z\right)  =\psi_{0}\left(  z+t\right)  ,\quad
\tilde{\psi}_{0,\kappa}^{t}\left(  z\right)  =\tilde{\psi}_{0}\left(
z-t\right)
\]
is extended by ultra relativistic limit $\kappa\longrightarrow\infty$ from the
orthogonal Hardy classes $\mathcal{E}_{0}^{\mp}$ onto the inductive spaces
$\mathcal{G}_{0}^{\mp}$. By continuity they are uniquely defined as the
opposite plane propagations on the whole Hilbert space $\mathfrak{h}%
_{0}\otimes L^{2}\left(  \mathbb{R}\right)  $ where they satisfy the
connection $\tilde{\psi}_{0}\left(  -z\right)  =\sigma_{0}\psi_{0}\left(
z\right)  $.

In the second case $k+\omega_{\kappa}\left(  -k\right)  \leq m^{2}/2\varkappa$
for all$\quad\varkappa=\kappa-\kappa^{\circ}>\left\vert \mu\right\vert $ and
$k<\kappa^{\circ}$. Using the inequality $\left\vert e^{x}-1\right\vert
<2\left\vert x\right\vert $ for any $x\in\mathbb{C}$ with $\left\vert
x\right\vert \leq1$ we obtain the estimate
\[
\left\Vert I\left(  \kappa^{\circ},\kappa\right)  \right\Vert \leq\left\Vert
e^{-\mathrm{i}\left(  k+\omega_{\kappa}\left(  -k\right)  \right)
t}-1\right\Vert <2\left\vert t\right\vert \left\Vert k+\omega_{\kappa}\left(
-k\right)  \right\Vert <\left\vert t\right\vert \frac{m^{2}}{\varkappa}%
\]
for the integral $I\left(  \kappa^{\circ},\kappa\right)  $ with $\left\Vert
g\right\Vert ^{2}=\frac{1}{2\pi}\int\left\Vert g\left(  k\right)  \right\Vert
^{2}\mathrm{d}k\leq1$. Hence for any $\kappa^{\circ}>0$, $\varepsilon>0$ and
each $t\in\mathbb{R}$ there exists a $\kappa^{\prime}<\infty$ such that
$\left\Vert I\left(  \kappa^{\circ},\kappa\right)  \right\Vert <\varepsilon$
for all $\kappa>\kappa^{\prime}$. Namely, one can take $\kappa^{\prime}%
=\kappa^{\circ}+\max\left\{  m,\left\vert t\right\vert m^{2}/\varepsilon
\right\}  $ such that $\varkappa=\kappa-\kappa^{\circ}>\kappa^{\prime}%
-\kappa>m$ and $\left\vert t\right\vert m^{2}/\varkappa<\varepsilon$. Thus the
plane wave propagation is indeed the ultra relativistic limit of (\ref{3.9})
in the inductive uniform sense.

In the third case one should replace $\mathfrak{h}=L^{2}\left(  \mathbb{R}%
^{d}\right)  $ by the inductive limit $\mathfrak{h}^{\circ}=\cup
\mathfrak{h}_{\mathbf{\kappa}}$ of Hilbert subspaces $\mathfrak{h}%
_{\mathbf{\kappa}}$ of functions in $L^{2}\left(  \mathbb{R}^{d}\right)  $
having the localized Fourier amplitudes $h\left(  \mathbf{k}\right)  =0$,
$\mathbf{k}\notin\left(  -\mathbf{\kappa},\mathbf{\kappa}\right)  $ for a
$\mathbf{\kappa\in}\mathbb{R}^{d}$. Then $\mu_{0}^{2}-\nabla^{2}<\mu_{0}%
^{2}+\mathbf{\kappa}^{2}$ in each $\mathfrak{h}_{\mathbf{\kappa}}$, and
$\left\|  I\left(  \kappa^{\circ},\kappa\right)  \right\|  <\left|  t\right|
\left(  \mu_{0}^{2}+\mathbf{\kappa}^{2}\right)  /\varkappa$ if $\left\|
g\right\|  \leq1$ for the Fourier amplitudes of $\rho_{0}^{1/2}\psi_{0}%
\in\mathfrak{h}_{\mathbf{\kappa}}\otimes\mathcal{E}_{\kappa^{\circ}}^{-}$ and
of $\rho_{0}^{1/2}\tilde{\psi}_{0}\in\mathfrak{h}_{\mathbf{\kappa}}%
\otimes\mathcal{E}_{\kappa^{\circ}}^{+}$. Hence for any $\kappa^{\circ}>0$,
$\mathbf{\kappa\in}\mathbb{R}^{d}$, $\varepsilon>0$ and each $t\in\mathbb{R}$
there exists a $\kappa^{\prime}<\infty$ such that $\left\|  I\left(
\kappa^{\circ},\kappa\right)  \right\|  <\varepsilon$ for all $\kappa
>\kappa^{\prime}$, namely
\[
\kappa^{\prime}=\kappa^{\circ}+\max\left\{  \sqrt{\mu_{0}^{2}+\mathbf{\kappa
}^{2}},\left|  t\right|  \left(  \mu_{0}^{2}+\mathbf{\kappa}^{2}\right)
/\varepsilon\right\}  .
\]
However the estimate $\left|  t\right|  \left(  \mu_{0}^{2}+\mathbf{\kappa
}^{2}\right)  /\left(  \kappa-\kappa^{\circ}\right)  $ depends now on
$\mathbf{\kappa}$ defining the choice of $g\left(  k\right)  $ in
$\mathfrak{h}^{\circ}$ for each $k<\kappa^{\circ}$. This proves that the plane
wave propagation is the ultra relativistic limit of (\ref{3.9}) also in the
general vector case, although not in the uniform but in the strong inductive
convergence sense.

Thus the boundary value problem (\ref{3.7}) in the ultra relativistic limit is
unitary equivalent to the plane propagations (\ref{3.5}) of opposite waves
$\psi_{0},\tilde{\psi}_{0}$ with the connection $\tilde{\psi}_{0}\left(
-z\right)  =\sigma_{0}\psi_{0}\left(  z\right)  $ for all $z\in\mathbb{R}$. In
the half space $z\in\mathbb{R}^{+}$ this obviously can be written as
\[
\psi_{0}^{t}\left(  z\right)  =\chi_{0}^{t}\left(  z\right)  ,z\geq
0,\quad\text{ }\tilde{\psi}_{0}^{t}\left(  -z\right)  =\chi_{0}^{t}\left(
z_{-}\right)  ,z\leq0,
\]
where $\chi_{0}^{t}\left(  z\right)  =\chi_{0,t}\left(  z+t\right)  $ is the
truncated input wave (\ref{3.6}) with $\psi_{0},\sigma_{0}$ instead of
$\psi,\sigma$. Returning back to $\psi^{t}=\hat{\epsilon}_{\varkappa}^{\ast
}\psi_{0}^{t}$ and $\tilde{\psi}^{t}=\check{\epsilon}_{\varkappa}^{\ast}%
\tilde{\psi}_{0}^{t}$ we shall obtain the representation (\ref{3.8}) with
\[
\chi^{t}\left(  z\right)  =\epsilon_{\varkappa}^{\ast}\left(  z\right)
e^{t\partial_{z}}\epsilon_{\varkappa}\left(  z\right)  \chi_{t}\left(
z\right)  =\epsilon_{\varkappa}\left(  z,z+t\right)  \chi_{t}\left(
z+t\right)  ,
\]
due to continuity of $\epsilon_{\varkappa}\left(  z,z+t\right)  $, where
$\chi_{t}=\hat{\epsilon}_{\varkappa}^{\ast}\chi_{0,t}$ is given in (\ref{3.8}).
\end{proof}

\begin{remark}
Let the operator $\varkappa_{0}=\varkappa\left(  0\right)  $ be self-adjoint
in $\mathfrak{h}$, $e^{-\mathrm{i}\varkappa_{0}t}$ be the one-parametric group
generated by $\varkappa_{0}$ in $\mathfrak{h}_{0}$, and $\epsilon_{\upsilon
}\left(  z\right)  =e^{\mathrm{i}\varkappa_{0}z}\epsilon_{\varkappa}\left(
z\right)  $ be the exponential family (\ref{2.5}) with the generator
\[
\upsilon\left(  z\right)  =\varkappa\left(  z\right)  -\epsilon_{\varkappa
}^{\ast}\left(  z\right)  \varkappa_{0}\epsilon_{\varkappa}\left(  z\right)  .
\]
Then the truncated wave $\chi^{t}=\hat{1}_{0}^{\bot}\psi^{t}+\hat{1}_{0}%
\tilde{\psi}^{t}$ in the interaction representation $\chi\left(  t\right)
=e^{\mathrm{i}\hat{\gamma}t}\chi^{t}$ with respect to the unitary group
generated by $\hat{\gamma}=\hat{\upsilon}+\mathrm{i}\partial_{z}$ satisfies
the stochastic single-jump equation
\begin{equation}
\mathrm{d}\chi\left(  t,z\right)  +\mathrm{i}\varkappa_{\upsilon}\left(
z\right)  \chi\left(  t,z\right)  \mathrm{d}t=\left(  \sigma_{\upsilon}\left(
z\right)  -1\right)  \chi\left(  t,z\right)  \mathrm{d}1_{t}\left(  z\right)
,\;t>0\label{3.10}%
\end{equation}
with the $e^{-\mathrm{i}\hat{\gamma}t}$-invariant density $\rho\left(
z\right)  =\epsilon_{\upsilon}^{\ast}\left(  z\right)  \rho_{0}\epsilon
_{\upsilon}\left(  z\right)  $ and
\[
\varkappa_{\upsilon}\left(  z\right)  =\epsilon_{\upsilon}^{\ast}\left(
z\right)  \varkappa_{0}\epsilon_{\upsilon}\left(  z\right)  ,\quad
\sigma_{\upsilon}\left(  z\right)  =\epsilon_{\upsilon}^{\ast}\left(
z\right)  \sigma_{0}\epsilon_{\upsilon}\left(  z\right)  .
\]

\end{remark}

Indeed, as the density function $\rho\left(  z\right)  $ is symmetric and
invertible,
\[
\varkappa\left(  z\right)  -\varkappa^{\ast}\left(  z\right)  =\mathrm{i}%
\rho\left(  z\right)  ^{-1}\partial_{z}\rho\left(  z\right)  =\mathrm{i}%
\tilde{\rho}\left(  z\right)  ^{-1}\tilde{\partial}_{z}\tilde{\rho}\left(
z\right)  =\tilde{\varkappa}^{\ast}\left(  z\right)  -\tilde{\varkappa}\left(
z\right)  ,
\]
and thus $\left(  \varkappa+\tilde{\varkappa}\right)  ^{\ast}=\varkappa
+\tilde{\varkappa}$. In particular, the operator $\varkappa_{0}=\varkappa
\left(  0\right)  $ is Hermitian in $\mathfrak{h}_{0}$, $\varkappa_{0}^{\ast
}=\rho_{0}^{-1}\varkappa_{0}^{+}\rho_{0}=\varkappa_{0}$, and it is
self-adjoint in $\mathfrak{h}_{0}$, commuting with $\rho_{0}$ due to
self-adjointness $\varkappa_{0}^{+}=\varkappa_{0}$ in $\mathfrak{h}$ and
invertibility of $\rho_{0}$. Thus the dynamical group $e^{-\mathrm{i}%
\varkappa_{0}t}$ is unitary in $\mathfrak{h}_{0}$. It defines the unitary
exponential family $\hat{\epsilon}_{\upsilon}=\hat{\epsilon}_{\varkappa_{0}%
}^{\ast}\hat{\epsilon}_{\varkappa}$ from $L_{\mathfrak{h}}^{2}\left(
\mathbb{R},\rho\right)  $ onto $\mathfrak{h}_{0}\otimes L^{2}\left(
\mathbb{R}\right)  $ satisfying the equation
\[
\mathrm{i}\partial_{z}\epsilon_{\upsilon}\left(  z\right)  =\epsilon
_{\upsilon}\left(  z\right)  \varkappa\left(  z\right)  -\varkappa_{0}%
\epsilon_{\upsilon}\left(  z\right)  =\epsilon_{\upsilon}\left(  z\right)
\left(  \varkappa\left(  z\right)  -\varkappa_{\upsilon}\left(  z\right)
\right)  =\epsilon_{\upsilon}\left(  z\right)  \upsilon\left(  z\right)
\]
as $\epsilon_{\upsilon}^{\ast}\left(  z\right)  \varkappa_{0}\epsilon
_{\upsilon}\left(  z\right)  =\epsilon_{\varkappa}^{\ast}\left(  z\right)
\varkappa_{0}\epsilon_{\varkappa}\left(  z\right)  $ due to commutativity of
$e^{\mathrm{i}\varkappa_{0}t}$ and $\varkappa_{0}$. The one-parametric group
$\epsilon_{\upsilon}\left(  z,z+t\right)  e^{t\partial_{z}}$ is apparently
generated by the operator $\gamma\left(  z,\mathrm{i}\partial_{z}\right)
=\upsilon\left(  z\right)  +\mathrm{i}\partial_{z}$ which is the symbol of
generator $\hat{\gamma}$ for the unitary group evolution $e^{-\mathrm{i}%
t\hat{\gamma}}$. It is a unitary group in $L_{\mathfrak{h}}^{2}\left(
\mathbb{R},\rho\right)  $ as
\[
e^{\mathrm{i}t\hat{\gamma}}\rho e^{-\mathrm{i}t\hat{\gamma}}=\epsilon
_{\upsilon}^{\ast}e^{t\partial_{z}}\epsilon_{\varkappa_{0}}^{\ast}\rho
_{0}\epsilon_{\varkappa_{0}}e^{-t\partial_{z}}\epsilon_{\upsilon}%
=\epsilon_{\upsilon}^{\ast}\rho_{\varkappa_{0}}\epsilon_{\upsilon}=\rho
\]
due to the $z$-independence of $\rho_{\varkappa_{0}}=\epsilon_{\varkappa_{0}%
}^{\ast}\rho_{0}\epsilon_{\varkappa_{0}}=\rho_{0}$. Hence the truncated wave
in the interaction representation is given by
\begin{align*}
\chi\left(  t,z\right)   &  =\epsilon_{\upsilon}\left(  z,z-t\right)  \chi
^{t}\left(  z-t\right)  =\epsilon_{\upsilon}\left(  z,z-t\right)
\epsilon_{\varkappa}\left(  z-t,z\right)  \chi_{t}\left(  z\right)  \\
&  =\epsilon_{\upsilon}^{\ast}\left(  z\right)  \epsilon_{\upsilon}\left(
z-t\right)  \epsilon_{\varkappa}^{\ast}\left(  z-t\right)  \epsilon
_{\varkappa}\left(  z\right)  \chi_{t}\left(  z\right)  =\epsilon_{\upsilon
}^{\ast}\left(  z\right)  e^{\mathrm{i}\left(  z-t\right)  \varkappa_{0}}%
\chi_{0,t}\left(  z\right)  ,
\end{align*}
where $\chi_{0,t}=\chi_{0}+\left(  \sigma_{0}-1\right)  \left(  1_{t}%
-1_{0}\right)  \chi_{0}$ with $\chi_{0}=\hat{1}_{0}^{\bot}\psi_{0}+\hat{1}%
_{0}\tilde{\psi}_{0}$. Taking into account that $\mathrm{d}t\mathrm{d}%
1_{t}\left(  z\right)  =0$ in the Hilbert space sense as it is zero almost
everywhere due to $\mathrm{d}1_{t}\left(  z\right)  =1\gg\mathrm{d}t\neq0$
only for the single point $z=t$ having the zero measure, we obtain
\begin{align*}
\mathrm{d}\chi\left(  t,z\right)   &  =\epsilon_{\upsilon}^{\ast}\left(
z\right)  e^{\mathrm{i}\left(  z-t\right)  \varkappa_{0}}\left[  \left(
\sigma_{0}-1\right)  \mathrm{d}1_{t}\left(  z\right)  \chi_{0}\left(
z\right)  -\mathrm{i}\varkappa_{0}\chi_{0,t}\left(  z\right)  \mathrm{d}%
t\right]  \\
&  =\epsilon_{\upsilon}^{\ast}\left(  z\right)  \left[  \left(  \sigma
_{0}-1\right)  \mathrm{d}1_{t}\left(  z\right)  -\mathrm{i}\epsilon_{\upsilon
}^{\ast}\left(  z\right)  \varkappa_{0}\mathrm{d}t\right]  e^{\mathrm{i}%
\left(  z-t\right)  \varkappa_{0}}\chi_{0,t}\left(  z\right)  \\
&  =\left[  \left(  \epsilon_{\upsilon}^{\ast}\left(  z\right)  \sigma
_{0}\epsilon_{\upsilon}\left(  z\right)  -1\right)  \mathrm{d}1_{t}\left(
z\right)  -\mathrm{i}\epsilon_{\upsilon}^{\ast}\left(  z\right)  \varkappa
_{0}\epsilon_{\upsilon}\left(  z\right)  \mathrm{d}t\right]  \chi\left(
t,z\right)  .
\end{align*}
Here we used that $\mathrm{d}1_{t}\left(  z\right)  =\mathrm{d}1_{0}\left(
z-t\right)  =0$ if $z\neq t$ such that
\[
\mathrm{d}1_{t}\left(  z\right)  e^{\mathrm{i}\left(  z-t\right)  \varkappa
}\chi_{0,t}\left(  z\right)  =\mathrm{d}1_{t}\left(  z\right)  \chi
_{0,t}\left(  z\right)  =\mathrm{d}1_{t}\left(  z\right)  \chi_{0,z}\left(
z\right)
\]
due to $\chi_{0,t}\left(  z\right)  |_{t=z}=\chi_{0}\left(  z\right)  $ as
$1_{t}\left(  z\right)  -1_{0}\left(  z\right)  =0$ for any $z\geq t\geq0$.
Thus we have proved that $\chi\left(  t,z\right)  $ indeed satisfies the
stochastic single jump equation (\ref{3.10}) in the Hilbert space
$L_{\mathfrak{h}}^{2}\left(  \mathbb{R},\rho\right)  $ of the initial
conditions $\chi=\hat{1}_{0}^{\bot}\psi+\hat{1}_{0}\tilde{\psi}$ with respect
to the unitary group evolution $e^{-\mathrm{i}t\left(  \hat{\upsilon
}+\mathrm{i}\partial_{z}\right)  }\left(  z,\mathrm{i}\partial_{z}\right)
=\epsilon_{\upsilon}\left(  z,z+t\right)  e^{t\partial_{z}}$.

In the particular case of the scalar-valued density $\rho\left(  z\right)  $
and $\upsilon\left(  z\right)  =\varkappa\left(  z\right)  -\varkappa
_{0}=\hbar^{-1}u\left(  z\right)  $ with $\varkappa_{0}=\hbar^{-1}H$,
$\sigma_{0}=S$ in the Hilbert space $\mathfrak{h}_{0}=\mathcal{H}$ we obtain
the stochastic equation (\ref{1.2}) for the unitary cocycle $V\left(
t,s\right)  =e^{-t\partial_{z}}V^{t}$, where $V^{t}=S^{\hat{1}_{0}}e^{t\left(
\partial_{z}-\mathrm{i}\hbar^{-1}H\right)  }S^{-\hat{1}_{0}}$, as a
quantum-mechanical stochastic approximation. Namely, the toy Hamiltonian model
for the interpretation of discontinuous stochastic evolution in terms of the
strongly continuous unitary group resolving the Dirac boundary value problem
in extra dimension, is indeed the ultra relativistic inductive limit of a
Schr\"{o}dinger boundary-value problem with bounded from below Hamiltonian
$H_{\kappa}\left(  z,p\right)  =\hbar\omega_{\kappa,\varkappa}\left(
z,-\hbar^{-1}p\right)  $ with $\hbar\varkappa\left(  z\right)  =u\left(
z\right)  +H$.

Although the proof of inductive convergence for the boundary value problem
(\ref{3.7}) implies the ultra relativistic limit for the truncated waves
\[
\chi_{\kappa}^{t}=\hat{1}_{0}^{\bot}\psi_{\kappa}^{t}+\hat{1}_{0}\tilde{\psi
}_{\kappa}^{t}=\sigma_{\varkappa}^{\hat{1}_{0}}\psi_{\kappa}^{t},
\]
it is interesting to see how the unitary cocycle $v_{\kappa}\left(  t\right)
=e^{\mathrm{i}t\hat{\omega}_{\upsilon,\kappa}}v_{\kappa}^{t}$ of the
interaction representation for the truncated unitary group converges to the
stochastic cocycle $v\left(  t\right)  =e^{\mathrm{i}t\left(  \hat{\upsilon
}+\mathrm{i}\partial_{z}\right)  }v^{t}$, resolving the equation (\ref{3.10}).
As the truncated groups
\[
v_{\kappa}^{t}=\hat{\sigma}_{\varkappa}^{\hat{1}_{0}}e^{-\mathrm{i}%
t\hat{\omega}_{\varkappa,\kappa}}\hat{\sigma}_{\varkappa}^{-\hat{1}_{0}},\quad
v^{t}=\hat{\sigma}_{\varkappa}^{\hat{1}_{0}}e^{-\mathrm{i}t\left(
\hat{\varkappa}+\mathrm{i}\partial_{z}\right)  }\hat{\sigma}_{\varkappa
}^{-\hat{1}_{0}}%
\]
in $L_{\mathfrak{h}}^{2}\left(  \mathbb{R},\rho\right)  $ are unitary
equivalent to the groups
\[
\hat{v}_{\kappa}^{t}=\hat{\sigma}^{\hat{1}_{0}}e^{-\mathrm{i}t\hat{\omega
}_{\varkappa_{0},\kappa}}\hat{\sigma}^{-\hat{1}_{0}},\quad\hat{v}^{t}%
=\hat{\sigma}^{\hat{1}_{0}}e^{-\mathrm{i}t\left(  \varkappa_{0}+\mathrm{i}%
\partial_{z}\right)  }\hat{\sigma}^{-\hat{1}_{0}}%
\]
in $L_{\mathfrak{h}}^{2}\left(  \mathbb{R}\right)  $ by $\sqrt{\rho_{0}}%
\hat{\epsilon}_{\upsilon}$, where $\hat{\sigma}_{0}$ is the multiplication by
$\sigma\left(  z\right)  =e^{\mathrm{i}\varkappa_{0}z}\sigma e^{-\mathrm{i}%
\varkappa_{0}z}$, one can demonstrate this on the convergence of the cocycle
\[
\hat{v}_{\kappa}\left(  t\right)  =e^{\mathrm{i}t\hat{\omega}_{\kappa}}%
\hat{\sigma}^{\hat{1}_{0}}e^{-\mathrm{i}t\hat{\omega}_{\varkappa_{0},\kappa}%
}\hat{\sigma}^{-\hat{1}_{0}}=e^{-\mathrm{i}t\hat{\varepsilon}_{\kappa
,\kappa+\varkappa_{0}}}\left(  I+\hat{\pi}_{\kappa}^{t}\left(  \hat{\sigma
}-1\right)  \right)  \hat{\sigma}^{-\hat{1}_{0}}%
\]
to $\hat{v}\left(  t\right)  =e^{-\mathrm{i}t\varkappa_{0}}\left(  I+\hat
{1}_{t}\left(  \hat{\sigma}-1\right)  \right)  \hat{\sigma}^{-\hat{1}_{0}}$.
Here $\hat{\pi}_{\kappa}^{t}$ is the orthoprojector (\ref{3.4}) which should
converge to $\hat{1}_{t}$, and
\[
e^{-\mathrm{i}t\hat{\varepsilon}_{\kappa,\kappa+\varkappa_{0}}}=e^{\mathrm{i}%
t\hat{\omega}_{\kappa}}e^{\mathrm{i}\varkappa_{0}z}e^{-\mathrm{i}t\hat{\omega
}_{\kappa}}e^{-\mathrm{i}\varkappa_{0}z}=e^{\mathrm{i}t\hat{\varepsilon
}_{\kappa}}e^{-\mathrm{i}t\hat{\varepsilon}_{\kappa+\varkappa_{0}}}%
\]
should converge to the unitary evolution group $e^{-\mathrm{i}t\varkappa_{0}}$.

Let us show that this indeed takes place in the inductive convergence sense on
the example of one-dimensional massless kinetic energy $\varepsilon\left(
\kappa\right)  =\left\vert \kappa\right\vert $ assuming that the operator
$\varkappa_{0}$ is bounded from below. In this case the generator
\[
\hat{\varepsilon}_{\kappa,\kappa+\varkappa_{0}}=\varepsilon\left(
\kappa+\varkappa_{0}+\mathrm{i}\partial_{z}\right)  -\varepsilon\left(
\kappa+\mathrm{i}\partial_{z}\right)
\]
of the dynamical group $e^{-\mathrm{i}t\hat{\varepsilon}_{\kappa
,\kappa+\varkappa_{0}}}$ converges trivially to $\varkappa_{0}$ in the
inductive space $\mathcal{G}^{-}$ as for any $\kappa^{\circ}>0$ there exists
such $\kappa^{\prime}<\infty$ that
\[
\left(  \left\vert \kappa+\varkappa_{0}+\mathrm{i}\partial_{z}\right\vert
-\left\vert \kappa+\mathrm{i}\partial_{z}\right\vert \right)  e^{\mathrm{i}%
kz}=\left(  \left\vert \kappa+\varkappa_{0}-k\right\vert -\left\vert
\kappa-k\right\vert \right)  e^{\mathrm{i}kz}%
\]
is equal $\varkappa_{0}e^{\mathrm{i}kz}$ for all $k<\kappa^{\circ}$. One can
take $\kappa^{\prime}=\kappa^{\circ}+\kappa_{0}$, where $-\kappa_{0}%
\leq\varkappa_{0}$ is the lower bound for $\varkappa_{0}$ such that
$e^{-\mathrm{i}t\hat{\varepsilon}_{\kappa,\kappa+\varkappa_{0}}}%
\psi=e^{-\mathrm{i}t\varkappa_{0}}\psi$ for all $\kappa\geq\kappa^{\prime}$ if
$\psi\in\mathcal{E}_{\kappa^{\circ}}^{-}$.

In order to prove the inductive convergence $\hat{\pi}_{\kappa}^{t}%
\longrightarrow\hat{1}_{t}$ we shall use the Fourier integral representation
\[
e^{-\tau\varepsilon\left(  \kappa\right)  }=\int_{-\infty}^{\infty
}e^{\mathrm{i}s\kappa}g_{\tau}\left(  s\right)  \mathrm{d}s,\quad g_{\tau
}\left(  s\right)  =\frac{1}{2\pi\mathrm{i}}\left(  \frac{1}{s-\mathrm{i}\tau
}-\frac{1}{s+\mathrm{i}\tau}\right)  ,
\]
where $\tau=\theta+\mathrm{i}t$, $\theta=\operatorname{Re}\tau>0$ is the
regularizing parameter for this generalized integral at $\tau=\mathrm{i}t$.

Indeed, the orthoprojector $\hat{\pi}_{\kappa}^{t}$ can be represented as
$\lim_{\theta\searrow0}\hat{\pi}_{\kappa}^{t,\theta}$of the pseudo
differential operator $\hat{\pi}_{\kappa}^{t,\theta}=e^{-\bar{\tau}%
\hat{\varepsilon}_{\kappa}}\hat{1}_{0}e^{-\tau\hat{\varepsilon}_{\kappa}}$
having the symbol
\begin{align*}
\pi_{\kappa}^{t,\theta}\left(  z,\mathrm{i}\partial_{z}\right)   &
=\int_{-\infty}^{0}\mathrm{d}r\frac{1}{2\pi}\int_{-\infty}^{\infty}%
e^{-\bar{\tau}\varepsilon\left(  \kappa+\mathrm{i}\partial_{z}\right)
}e^{\mathrm{i}\varkappa\left(  r-z\right)  }e^{-\tau\varepsilon\left(
\kappa+\mathrm{i}\partial_{z}\right)  }\mathrm{d}\varkappa\\
&  =\int_{-\infty}^{0}\mathrm{d}r\frac{1}{2\pi}\int_{-\infty}^{\infty
}e^{\mathrm{i}\left(  r-z\right)  \varkappa}e^{-\bar{\tau}\varepsilon\left(
\kappa+\varkappa+\mathrm{i}\partial_{z}\right)  -\tau\varepsilon\left(
\kappa+\mathrm{i}\partial_{z}\right)  }\mathrm{d}\varkappa\\
&  \int_{z}^{\infty}e^{\mathrm{i}s\left(  \kappa+\mathrm{i}\partial
_{z}\right)  }\mathrm{d}s\frac{1}{2\pi}\int_{-\infty}^{\infty}e^{-\mathrm{i}%
s\varkappa}e^{-\bar{\tau}\varepsilon\left(  \varkappa\right)  -\tau
\varepsilon\left(  \kappa+\mathrm{i}\partial_{z}\right)  }\mathrm{d}%
\varkappa\\
&  =\int_{z}^{\infty}e^{\mathrm{i}s\left(  \kappa+\mathrm{i}\partial
_{z}\right)  }\overline{g_{\tau}}\left(  s\right)  \mathrm{d}se^{-\tau
\varepsilon\left(  \kappa+\mathrm{i}\partial_{z}\right)  }=e_{\kappa}%
^{\bar{\tau}}\left(  z,\mathrm{i}\partial_{z}\right)  e^{-\tau\hat
{\varepsilon}_{\kappa}}.
\end{align*}
Here $\overline{g_{\tau}}\left(  s\right)  =g_{\bar{\tau}}\left(  -s\right)
=g_{\bar{\tau}}\left(  s\right)  $ defines the operator $e^{-\bar{\tau}%
\hat{\varepsilon}_{\kappa}}$ as the limit of the contour integral
\[
e_{\kappa}^{\bar{\tau}}\left(  z,\mathrm{i}\partial_{z}\right)  =\frac{1}%
{2\pi\mathrm{i}}\int_{z}^{\infty}e^{\mathrm{i}s\left(  \kappa+\mathrm{i}%
\partial_{z}\right)  }\left(  \frac{1}{s-\mathrm{i}\bar{\tau}}-\frac
{1}{s+\mathrm{i}\bar{\tau}}\right)  \mathrm{d}s
\]
at $z\longrightarrow-\infty$. Let us show that this contour integral has the
same limit $e^{-\bar{\tau}\left(  \kappa-k\right)  }$ at $\kappa
\longrightarrow\infty$ for a finite $z<t$, and zero limit at $z\geq t$ in the
inductive sense. Indeed, for each $\kappa^{\circ}>0$ the contour integral
\[
e_{\kappa}^{\bar{\tau}}\left(  z,\mathrm{i}\partial_{z}\right)  e^{\mathrm{i}%
kz}=e^{\mathrm{i}kz}\frac{1}{2\pi\mathrm{i}}\int_{z}^{\infty}e^{\mathrm{i}%
s\left(  \kappa-k\right)  }\left(  \frac{1}{s-\mathrm{i}\bar{\tau}}-\frac
{1}{s+\mathrm{i}\bar{\tau}}\right)  \mathrm{d}s
\]
can be closed at $\kappa\gg\kappa^{\circ}$ for all $k<\kappa^{\circ}$ via the
upper plane of complex $s$ by extending the integration to the vertical from
$s=z+\mathrm{i}\infty$ to $z\in\mathbb{R}$ without essential change of its
value due to the exponential decay of $\left\vert e^{\mathrm{i}s\left(
\kappa-k\right)  }\right\vert <e^{-\operatorname{Im}s\left(  \kappa
-\kappa^{\circ}\right)  }$ for each $\operatorname{Im}s>0$. As the only pole
of the integrand in the upper plane is $s_{+}=t+\mathrm{i}\theta
=\mathrm{i}\bar{\tau}$ (the only other pole $s_{-}=-\mathrm{i}\bar{\tau}$ is
in the lower plane) the major value of this integral is defined by the
integrand value
\[
\frac{1}{2\pi\mathrm{i}}\left(  \int_{z+\mathrm{i}\infty}^{z}+\int_{z}%
^{\infty}\right)  e^{\mathrm{i}s\left(  \kappa-k\right)  }\frac{1}%
{s-\mathrm{i}\bar{\tau}}\mathrm{d}s=e^{-\bar{\tau}\left(  \kappa-k\right)
}1_{t}\left(  z\right)
\]
at the pole $s_{+}$ if it is inside the contour ($z<t$), and it is zero if
this pole is outside of the contour ($z\geq t$). Thus
\[
\pi_{\kappa}^{t,\theta}\left(  z,\mathrm{i}\partial_{z}\right)  e^{\mathrm{i}%
kz}\approx e^{\mathrm{i}kz}e^{-\bar{\tau}\left(  \kappa-k\right)  }%
1_{t}\left(  z\right)  e^{-\tau\left(  \kappa-k\right)  }=e^{\mathrm{i}%
kz}1_{t}\left(  z\right)  e^{-2\theta\left(  \kappa-k\right)  }%
\]
as $\kappa\gg k$ where the operator $e^{-\tau\hat{\varepsilon}_{\kappa}}$
coincides with $e^{-\tau\left(  \kappa-k\right)  }$:
\[
e^{-\tau\varepsilon\left(  \kappa+\mathrm{i}\partial_{z}\right)
}e^{\mathrm{i}kz}=e^{\mathrm{i}kz}e^{-\tau\left\vert \kappa-k\right\vert
}=e^{\mathrm{i}kz}e^{-\tau\left(  \kappa-k\right)  },\quad\forall
\kappa>k\text{.}%
\]
This proves the inductive convergence of $\hat{\pi}_{\kappa}^{t}=\lim
_{\theta\searrow0}\hat{\pi}_{\kappa}^{t,\theta}$ at the ultra relativistic
limit $\kappa\longrightarrow\infty$ towards the integrator $\hat{1}_{t}$ of
the stochastic equation (\ref{3.10}).

Thus the Hamiltonian boundary value problem in $\mathbb{R}^{+}$, corresponding
to free propagation of input and output waves in the opposite directions with
the unitary reflection at $z=0$ can be reduced at the ultra relativistic limit
to the propagation problem in $\mathbb{R}$ with the unitary transition from
$z=0$ to $z=0_{-}$ corresponding to a $\delta$-function potential on
$\mathbb{R}$. In the interaction representation it is described by the
stochastic equation
\[
\mathrm{d}\hat{v}\left(  t\right)  +\mathrm{i}\varkappa_{0}\hat{v}\left(
t\right)  \mathrm{d}t=\left(  \sigma-1\right)  \hat{v}\left(  t\right)
\mathrm{d}\hat{1}_{t},\quad t>0,\hat{v}\left(  0\right)  =I.
\]


\begin{thebibliography}{9}                                                                                                %
\bibitem {Bel95}V. P. Belavkin, ''A Dynamical Theory of Quantum Measurement
and Spontaneous Localization,'' \textit{Russian Journal of Mathematical
Physics}, \textbf{3}, No. 1, 3--23 (1995)

\bibitem {AAFL91}L., Accardi, R., Alicki, A. Frigerio, and Y. G. Lu, ''An
invitation to the weak coupling and low density limits,'' \textit{Quantum
Probability and Related Topics} \textbf{VI}, 3-61 (1991)

\bibitem {Bel96}V. P. Belavkin, ''A stochastic Hamiltonian approach for
quantum jumps, spontaneous localizations, and continuous trajectories,''
\textit{Quantum Semicalss. Opt.} \textbf{8}, 167--187 (1996)

\bibitem {Bel94}V. P. Belavkin, ''Nondemolition Principle of Quantum
Measurement Theory,'' \textit{Foundations of Physics,} \textbf{24}, No. 5,
685--714 (1994)

\bibitem {Che97}A. M. Chebotarev, ''The quantum stochastic equation is
equivalent to a symmetric boundary value problem for the Schr\"{o}dinger
Equation,'' \textit{Mathematical Notes,} \textbf{61}, No. 4, 510--518 (1997)

\bibitem {HP84}R. L. Hudson, and K. R. Parthasarathy, ''Quantum Ito's formula
and stochastic evolutions,'' \textit{Comm. Math. Phys.} \textbf{93}, No. 3,
301--323 (1984)
\end{thebibliography}
\end{document}